\documentclass[journal]{IEEEtran}

\usepackage{xcolor,soul,framed} 

\colorlet{shadecolor}{yellow}
\usepackage[pdftex]{graphicx}
\graphicspath{{../pdf/}{../jpeg/}}
\DeclareGraphicsExtensions{.pdf,.jpeg,.png}

\usepackage[cmex10]{amsmath}
\usepackage{array}
\usepackage{mdwmath}
\usepackage{mdwtab}
\usepackage{eqparbox}
\usepackage{url}

\usepackage[compress]{cite}
\usepackage{graphics}
\usepackage{subfig}
\usepackage{booktabs}
\usepackage[font={small}]{caption}
\usepackage{newtxtext,newtxmath}
\let\labelindent\relax
\usepackage{enumitem}

\usepackage{color}
\newcommand{\rev}[1]{{\color{black}#1}}

\begin{document}
\captionsetup[figure]{name={Fig.},labelsep=period,justification=raggedright,font={small}}
\title{OAcode: Overall Aesthetic 2D Barcode on Screen}

\author{Zehua~Ma, Xi~Yang, Han~Fang, Weiming~Zhang, and Nenghai~Yu

\thanks{Zehua Ma, Xi Yang, Weiming Zhang, and Nenghai Yu are with CAS Key Laboratory of Electromagnetic Space Information, University of Science and Technology of China, Hefei, 230026, China. Han Fang is with School of Computing, National University of Singapore, 117417, Singapore. Corresponding author: Weiming Zhang. (e-mail: mzh045@mail.ustc.edu.cn, zhangwm@ustc.edu.cn)
}

\thanks{This work was supported in part by the Natural Science Foundation of China under Grant 62072421, 62102386, 62002334, 62121002 and U20B2047, and Students' Innovation and Entrepreneurship Foundation of USTC under Grant XY2022X01CY.
}
}
\maketitle

\begin{abstract}
Nowadays, two-dimensional (2D) barcodes have been widely used in various domains. And a series of aesthetic 2D barcode schemes have been proposed to improve the visual quality and readability of 2D barcodes for better integration with marketing materials. Yet we believe that the existing aesthetic 2D barcode schemes are \emph{partially} aesthetic because they only beautify the data area but retain the position detection patterns with the black-white appearance of traditional 2D barcode schemes. Thus, in this paper, we propose the first \emph{overall} aesthetic 2D barcode scheme, called OAcode, in which the position detection pattern is canceled. Its detection process is based on the pre-designed symmetrical data area of OAcode, whose symmetry could be used as the calibration signal to restore the perspective transformation in the barcode scanning process. Moreover, an enhanced demodulation method is proposed to resist the lens distortion common in the camera-shooting process. The experimental results illustrate that when $5 \times 5cm$ OAcode is captured with a resolution of $720\times1280$ pixels, at the screen-camera distance of $10cm$ and the angle less or equal to $25^{\circ}$, OAcode has $100\%$ detection rate and $99.5\%$ demodulation accuracy. For $10 \times 10cm$ OAcode, it could be extracted by consumer-grade mobile phones at a distance of $90cm$ with around $90\%$ accuracy.
\end{abstract}

\section{Introduction}\label{introduction}

\IEEEPARstart{W}{ith} the increasing popularity of smart mobile devices and wireless network infrastructures \cite{device}, two-dimensional (2D) barcodes are used in a wide range of industries, from manufacturing and logistics to mobile marketing and business. \rev{Especially in mobile marketing, 2D barcodes play a significant role since 2008 \cite{qr_mobile_marketing,cata2013qr}. According to a research statistic from 2012 \cite{okazaki2012benchmarking}, print media, i.e., magazines, flyers, and newspapers, are the most frequently used media to present 2D barcodes, whose proportion is over 80\%. Yet in 2022 \cite{usage_in_2022}, over 39\% of 2D barcodes are scanned from screens, including TV and website on PC. People are increasingly preferring to scan 2D barcodes on screen because, on the one hand, the screen content grows explosively \cite{min2021screen}, and on the other hand, the restriction of physical contact during the COVID-19 pandemic and the implementation of the paperless office contribute to a rapid decline in the number of print media \cite{ma2022learning,sellen2003myth}. Moreover, as media for mobile marketing, the screen has some advantages over print ones, such as higher color accuracy and better maintainability. By attaching 2D barcodes to marketing-related materials and displaying them on various screens, potential customers can easily access the relevant information about the product by scanning these 2D barcodes with their smart mobile devices.}

\begin{figure}[t]
    \begin{center}
    \subfloat[Some aesthetic 2D barcodes.]{
        {\centering\includegraphics[width=1.6in]{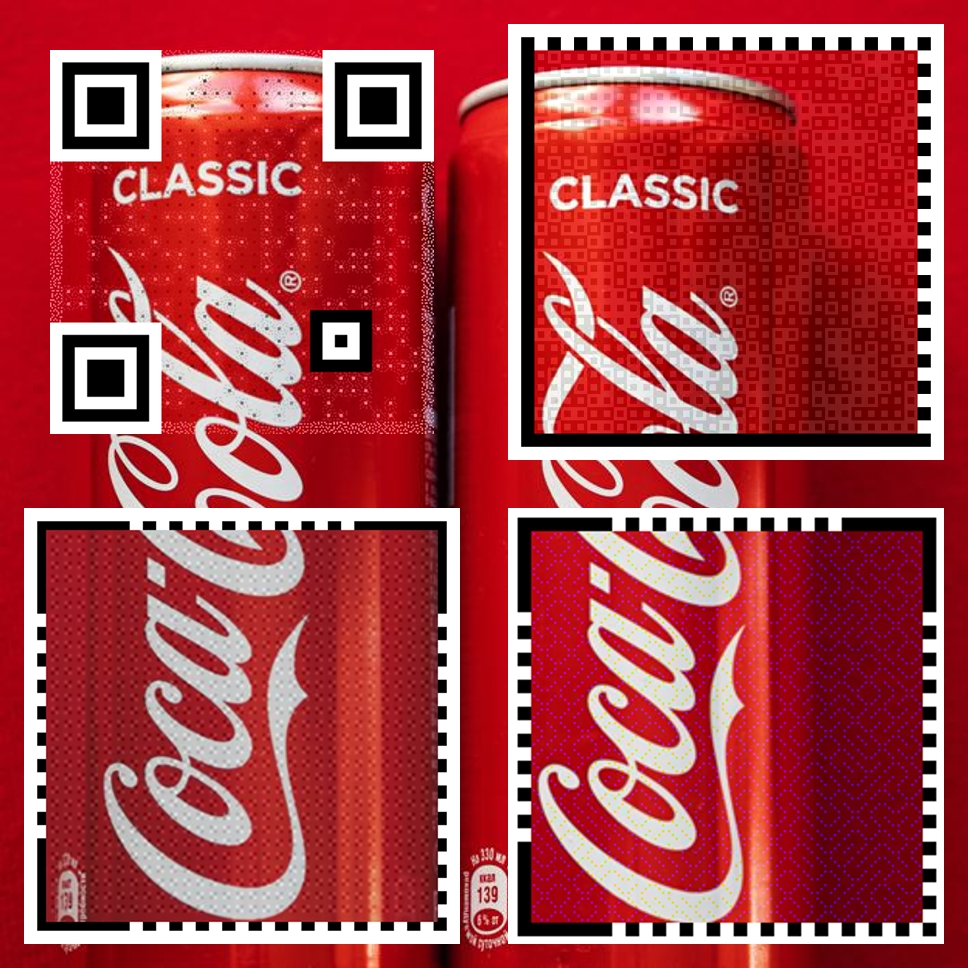}
        }
    }
    \subfloat[OAcode.]{
        {\centering\includegraphics[width=1.6in]{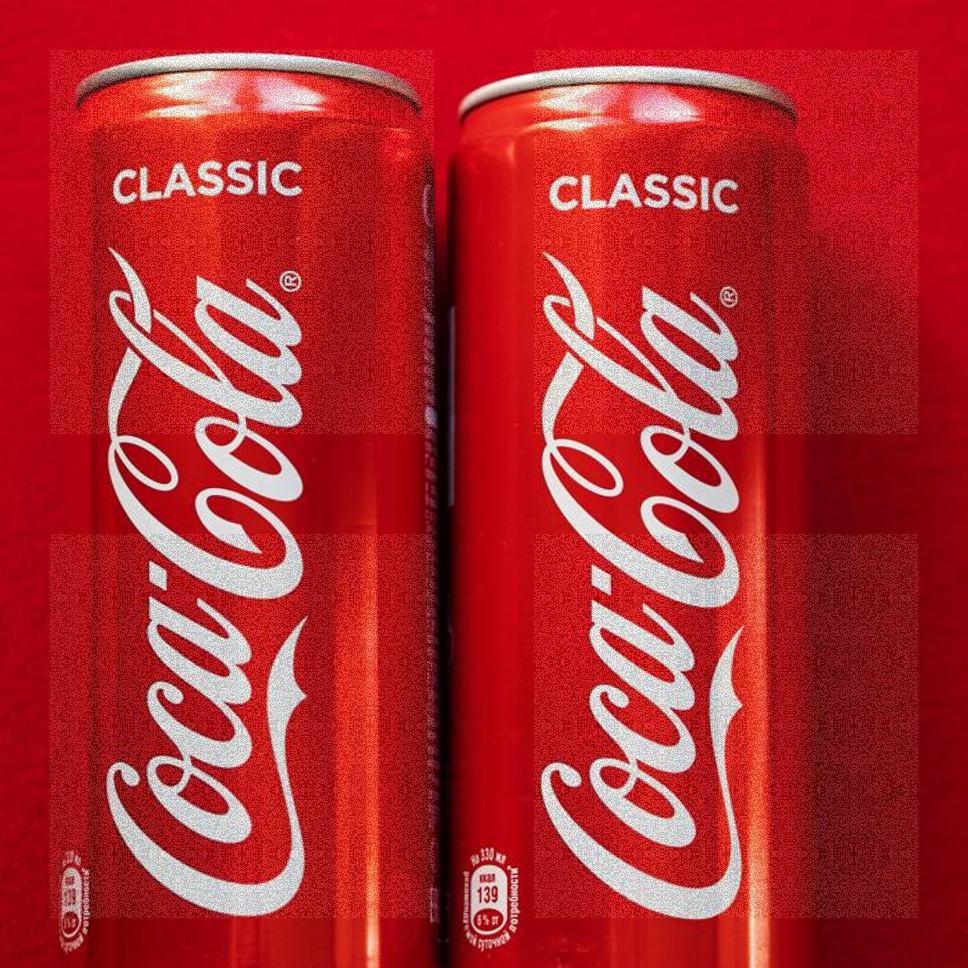}
        }
    }
    \end{center}
\caption{Example of a commercial poster embedded with aesthetic 2D barcodes. From top left to bottom right, (a) is embedded with QR image \cite{QRimage}, PiCode \cite{PiCode}, RA Code \cite{racode}, and RU code \cite{rucode}. Their beautified data areas make the semantics of the commercial poster could be preserved, but the black and white appearance of position detection patterns and the white quiet area cause a visual discontinuity in the poster. (b) embeds four OAcodes in the corresponding positions and have a better overall visual quality. }
\label{poster}
\end{figure}

However, traditional barcodes, like Quick Response (QR) code \cite{qrcode} and Data Matrix \cite{datamatrix}, have some drawbacks in mobile marketing. Their binary appearance is initially designed to be machine-friendly for fast data transmission, yet it lacks readable hints to humans. People cannot know what information is linked to a traditional 2D barcode before scanning it, which may limit the interest of potential consumers. \rev{Additionally, their black-and-white appearance is difficult to meet the increasing demand for visual quality \cite{zhai2020perceptual,min2017blind,min2018blind}, especially in mobile marketing on the screen \cite{min2021screen,min2020study,min2017unified,min2018saliency}. As a result, it is hard to integrate them with colorful and well-designed mobile marketing materials.}

\begin{figure}[t]
  \begin{center}
  \includegraphics[width=3.4in]{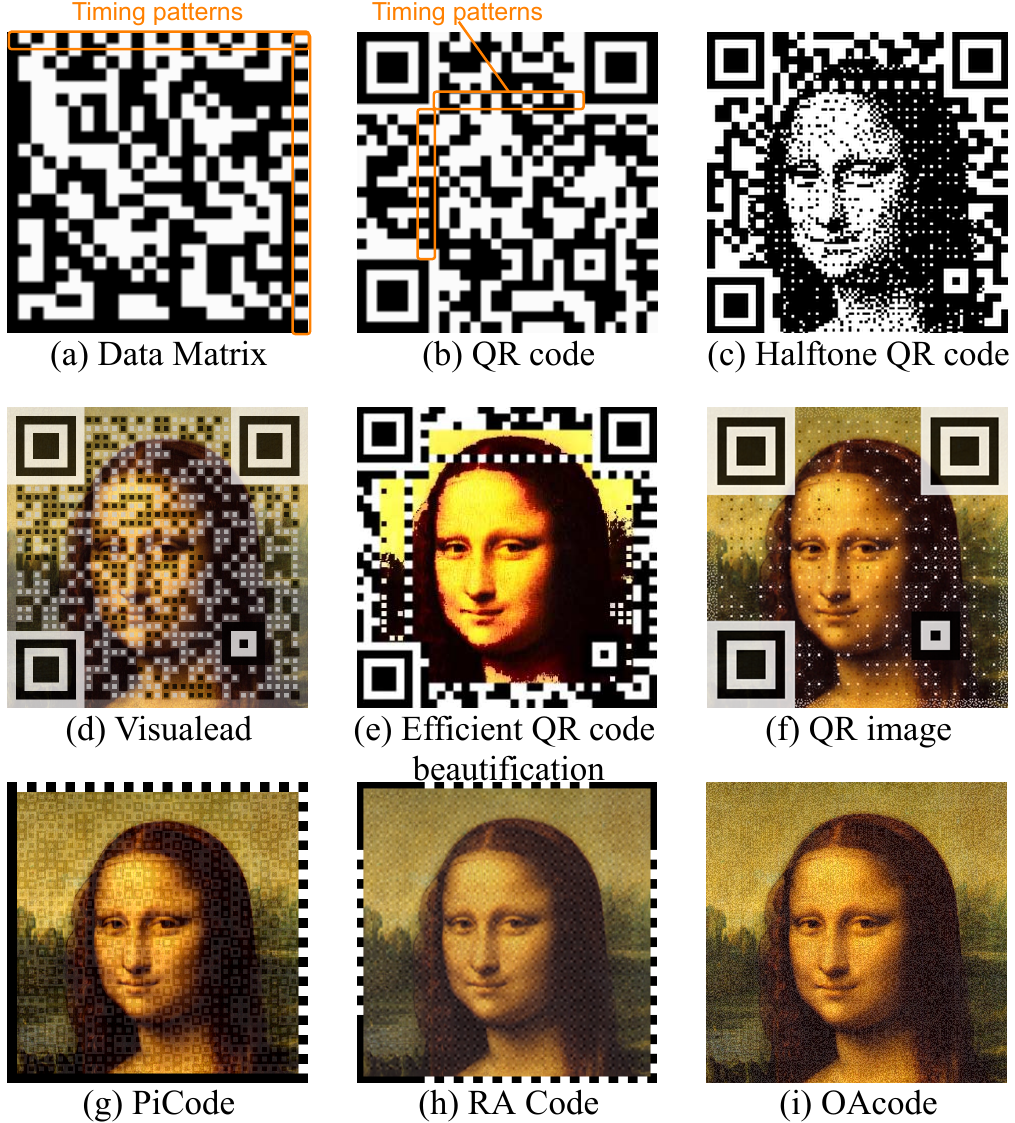}
  \end{center}
\caption{Some examples of 2D barcode schemes. From the top left to the bottom right, the shown 2D barcode is getting closer to the target picture. Except for the proposed OAcode, all 2D barcodes have obtrusive position detection patterns and timing patterns.}
\label{barcodes}
\end{figure}

To solve these problems, many types of aesthetic 2D barcode schemes are proposed, including the halftone 2D barcode \cite{qart,chu2013halftone}, the picture combination 2D barcode \cite{Eff_Beau}, and the picture-embedding \cite{Visualead,QRimage,PiCode,racode,rucode,Visually,appearance,Stylized,yang2016artcode,liu2011toward}, where the picture-embedding 2D barcode is considered the most promising technology \cite{Stylized}. It generally embeds a data-related picture in the data area to generate the aesthetic 2D barcode, making them more human-readable, attractive, and better integrated with marketing materials. As shown in Fig.~\ref{poster} (a), the main content of the commercial poster would not be obscured by these aesthetic 2D barcodes. \rev{Yet it could be founded that these existing schemes still retain obtrusive position detection patterns. Because in most 2D barcode detection schemes, even some state-of-art ones \cite{EMBDN,Tiny-BDN,ZhangMJZWZ21}, position detection patterns are essential. This design results in a low structural similarity of marketing materials before and after embedding the aesthetic 2D barcode, which is one of the most commonly used quality metrics in image quality assessment \cite{zhai2020perceptual,min2017blind,min2018blind}.} Additionally, as shown in Fig.~\ref{barcodes} (a) and (b) and Fig.~\ref{poster} (a), to ensure the position detection patterns keep the pre-designed appearance, white quiet zones should be set around them \cite{qr_area}. Position detection patterns and the quiet zone cause an apparent boundary between the aesthetic 2D barcode and the well-designed poster and undermine the aesthetics of the poster design. Thus, we believe that these existing aesthetic 2D barcodes are partially aesthetic.

Thus, in this paper, we propose an overall aesthetic 2D barcode, named as OAcode, which eliminates the visual impact of the position detection patterns while beautifying the data area. As shown in Fig.~\ref{poster} (b), the poster embedded with OAcode preserves the original design much better than existing aesthetic barcodes.

However, the position detection patterns play a significant role in the 2D barcode detection process with the following functions:

\noindent \textbf{Resynchronization} In the traditional 2D barcode detection process, four corner points of the 2D barcode could be detected based on position detection patterns, and then used to determine parameters to restore the perspective transformation. Moreover, position detection patterns are usually designed as rotationally invariant to determine the rotation state of the 2D barcode. Both of these are combined to resynchronize the scanned 2D barcodes to the unrotated square state.

\noindent \textbf{Parameters determination} To cope with the capacity requirements in different scenarios, a 2D barcode scheme usually has multiple capacity versions to choose from, which requires the extraction program to be able to determine the version of the scanned 2D barcode. In the existing traditional and aesthetic 2D barcode schemes, the version parameters are transmitted by the timing pattern, which is a part of position detection patterns as shown in Fig.~\ref{barcodes} (a) and (b). Besides the capacity version, the extraction program needs to know the module\footnote{According to the barcode terminology, the smallest square unit with the black-and-white appearance used to represent bit `0’ or `1’ is called a module.} size, which could also be obtained from the timing pattern.

\noindent \textbf{Lens distortion correction} The lens distortion introduced by photographing process, which results in a line-to-curve mapping \cite{lens_distortion}, would decrease the demodulation accuracy of 2D barcodes, especially in its large capacity version \cite{PiCode}. The existing 2D barcode schemes usually leverage timing patterns to resist lens distortions.

The difficulty of designing OAcode is that it, as one aesthetic 2D barcode scheme, should achieve these necessary functions mentioned above without position detection patterns. The novel aspects of OAcode, i.e., the main contributions of this paper, are summarized as follows.

\begin{enumerate}[leftmargin=*]
    \item OAcode cancels the position detection patterns and thus has a better visual quality. The elimination of position detection patterns is made possible by the designed symmetrical data area of OAcode, whose symmetry is used as synchronizing signal in the following detection process. In the OAcode generation process, we also propose a fast blending method with adaptive intensity, which can play a similar role to error diffusion dithering and avoid the clipped intensity due to the content of the background image.
    \item We carefully design an OAcode detection strategy based on its data area symmetry and Hough transform, which could implement all functions that should have been realized by visible position detection patterns, including 2D barcode synchronization, rotation state determination, module size determination, and information capacity determination.
    \item We propose an enhanced demodulation method for OAcode to obtain a higher demodulation accuracy under the lens distortion, which commonly exists in the camera-shooting process \cite{stirmark1,stirmark2,lens_distortion}. Specifically, the result of the first demodulation, i.e., the accumulated data unit would be divided into subunits to execute the seconding position detection process as reference signals and re-accumulate the detection results to obtain a new data unit with less lens distortion.
\end{enumerate}

\section{Related Work}\label{related work}
The proposed OAcode is a kind of aesthetic 2D barcode scheme, which embeds information into one image. Thus, in this section, we first review some related aesthetic 2D barcode schemes and then discuss the difference between aesthetic 2D barcodes and some information embedding techniques.

\subsection{Aesthetic 2D Barcode}
Compared with traditional 2D barcodes (Fig.~\ref{barcodes} (a) and (b)) with an obtrusive appearance, aesthetic 2D barcodes (Fig.~\ref{barcodes} (c)$\sim$(i)) are human-readable, more attractive, and can be better integrated with marketing materials. In addition, almost all aesthetic 2D barcode schemes have the same design, i.e., the combination of position detection patterns and beautified data area, in which their position detection patterns have a similar appearance and the same functionality as those of traditional 2D barcodes. For example, we could find that the structure of position detection patterns in Fig.~\ref{barcodes} (c)$\sim$(h) is similar to ones shown in Fig.~\ref{barcodes} (a) and (b).

The existing aesthetic 2D barcode schemes could be roughly divided into two categories according to their design purposes. 
The first category is generated by modifying an existing traditional 2D barcode scheme. These schemes are compatible with the standard extractor and have better visual quality. Because of the widespread usage of QR code, most schemes in this category are beautified QR code schemes \cite{Visualead, Visually,appearance,chu2013halftone,Eff_Beau,QRimage,Stylized}. Fig.~\ref{barcodes} (c)$\sim$(f) show some examples. These schemes usually modify the generation process of QR code, mainly the modulation process, to improve the visual quality. For compatibility, they retain the same extraction process as QR code, resulting in the same position detection patterns.

And another category includes the aesthetic 2D barcode schemes \cite{PiCode,racode,liu2011toward,yang2016artcode,rucode} generated independently of existing black-white 2D barcodes. Schemes in this category need not be compatible with existing 2D barcode extractors. Therefore they usually have more design freedom and better performance compared with schemes in the above category. As shown in Fig.~\ref{barcodes} (g) and (h), one obvious visual improvement of aesthetic 2D barcodes in this category is that they could design smaller position detection patterns, which usually means a closer appearance to the background image and better visual quality.

The proposed OAcode belongs to the second category, yet compared to the previous schemes, OAcode has a completely different design, i.e., the elimination of position detection patterns. As a result, OAcode is visually closer to one image (Fig.~\ref{barcodes} (i)) and can be better integrated with one poster (Fig.~\ref{poster} (b)).

\subsection{Information Embedding Techniques}\label{difference}
We discuss the following information embedding techniques, including watermark, steganography, and deep data hiding. Watermark and steganography are the two main branches of information hiding \cite{hiding}, in which digital image watermarking \cite{stirmark2,history2,fang2019camera,Fanghan} is frequently used to prove image ownership, designed with a focus on the visual imperceptibility and the robustness to various digital image distortions. Steganography \cite{steganography1,steganography2,steganography3} is used for covert communication, so the corresponding methods focus on the security and the capacity of the embedded secret message, in which the security usually means the hidden message is imperceptible to steganalysts. Deep data hiding \cite{2019stegastamp,TERA,RIHOOP,jia2022learning} is an emerging category of techniques that use neural networks to hide information in images and extract it through camera shooting. Deep data hiding focuses on the robustness of camera shooting distortions, which are more serious than digital image distortions. 

\rev{Although OAcode has a similar appearance to other information embedding techniques mentioned above, it follows the same design criteria as other aesthetic 2D barcode schemes, which makes OAcode different from these information embedding techniques. Specifically, one 2D barcode scheme always uses modules to represent bits `0' or `1'. For example, the modules of QR code are black-and-white square units and OAcode uses the spread spectrum module. Then, in the blending process, the background image only affects the blending intensity of these modules. Based on modules, 2D barcode schemes could design a parameter determination strategy, which allows its detector to extract one 2D barcode with unknown parameters, e.g., unknown capacity. Considering the different capacity and robustness requirements of 2D barcodes in various application scenarios, this strategy allows these 2D barcodes with different settings could be decoded by the same detector. While for most information embedding techniques mentioned above, the image content and the embedded information together determine the modification intensity and patterns on the host image to better balance the robustness and imperceptibility. Additionally, their end-to-end design has excellent performance with preset parameters but is not suitable for input with variable parameters.}

As summarized in Table~\ref{diff}, these four research areas are designed for different applications and have different concerns and design criteria.

\begin{table}[t]
\large
\caption{\\The different design criteria of information embedding techniques and aesthetic 2D barcode.} \centering
\label{diff}
\resizebox{\linewidth}{!}{
\begin{tabular}{c|ccc}
\toprule
\begin{array}{c}
     \textbf{Picture-embedding}  \\
     \textbf{methods}
\end{array} & \textbf{Robustness} & \textbf{Imperceptibility} &  \textbf{Capacity} \\
\midrule
\textbf{Steganography} & low & high & high \\
\midrule
\textbf{Watermarking} & medium & medium & low \\
\midrule
\textbf{Deep data hiding} & high & medium & medium \\
\midrule
\textbf{Aesthetic 2D barcode} & high & low & high \\
\bottomrule
\end{tabular}
}
\end{table}

\section{Symmetry Calculation}\label{sym_theory}
The proposed OAcode has a novel synchronization process, in which its symmetry is used as the calibration signal. To fast calculate symmetry, we slightly modify the method proposed by Ma et al. \cite{loc_distortion}. In this section, we first define the symmetry of a 2D matrix and then illustrate its calculation method. 
\subsubsection{Symmetry Definition}
If 2D matrix $\boldsymbol{A}$ is symmetrical about point $(x_s,y_s)$, $\boldsymbol{A}$ satisfies:
\begin{equation}
    A(x_s-x,y_s-y)=A(x_s+x,y_s+y).
\end{equation}
The point $(x_s,y_s)$ is one symmetrical center of $\boldsymbol{A}$. The symmetry $\boldsymbol{S}$ of $\boldsymbol{A}$ about a point $(i,j)$ can be defined by:
\begin{equation}\label{S1_def}
    S(i,j)=\sum_{x} \sum_{y} A(i-x,j-y)A(i+x,j+y).
\end{equation}
$S(i,j)$ is actually the correlation between $\boldsymbol{A}$ and the flipped $\boldsymbol{A}$ around point $(i,j)$. Meanwhile, given a point $(i,j)$, the value of $S(i,j)$ represents the possibility that point $(i,j)$ is one of the symmetrical centers of $\boldsymbol{A}$. In the ideal case, the positions of symmetrical peaks in $\boldsymbol{S}$ are the positions of symmetrical centers in $\boldsymbol{A}$.

\begin{figure*}[t]
  \begin{center}
  \includegraphics[width=\linewidth]{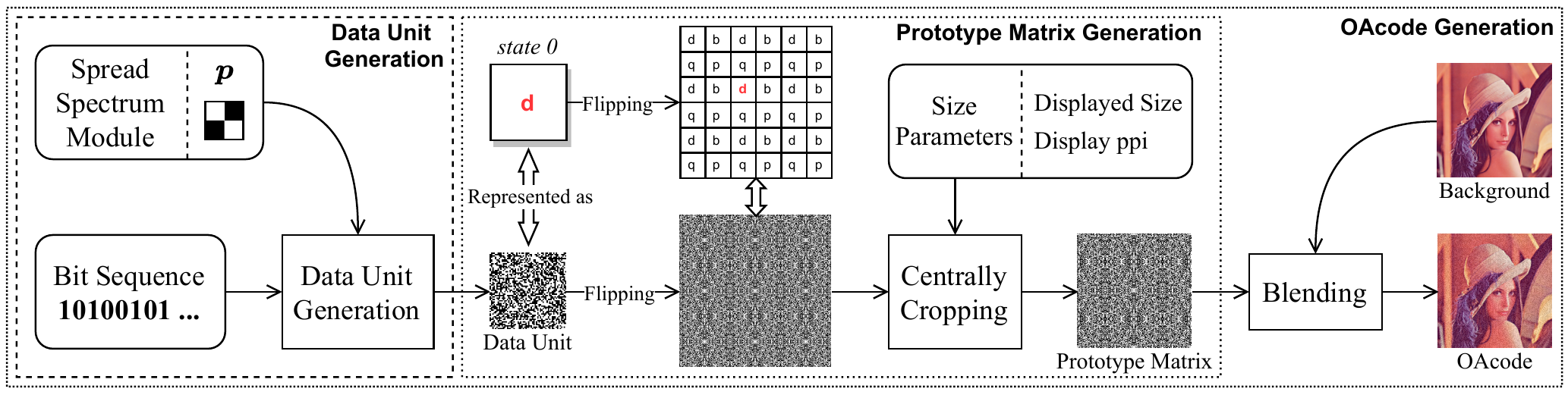}
  \end{center}
\caption{The generation process of OAcode, including data unit generation, prototype matrix generation, and blending process.}
\label{fig:generation}
\end{figure*}

\subsubsection{Auto-convolution Function}\label{sec:acnf}
Obviously, calculating symmetry in space domain like Eq.~(\ref{S1_def}) is inefficient, especially when the $\boldsymbol{A}$ has a large size.
To decrease the computational cost of Eq.~(\ref{S1_def}), it is first rewritten as:
\begin{equation}\label{S_def}
    \begin{aligned}
        S(i,j)&=\sum_{x} \sum_{y} A(i-x,j-y)A(i+x,j+y)\\
        &=\sum_{x} \sum_{y} A(x,y)A(2i-x,2j-y),
    \end{aligned}
\end{equation}
where $x$ and $y$ run over all values that lead to legal subscripts of $\boldsymbol{A}$.
Defining a temporary matrix $\boldsymbol{T}$, $T(2i,2j)=S(i,j)$ and substitute it to Eq.~(\ref{S_def}):
\begin{equation}\label{Temp_def}
    \begin{aligned}
        T(2i,2j)&=\sum_{x} \sum_{y} A(x,y)A(2i-x,2j-y)\\
        T(u,v)&=\sum_{x} \sum_{y} A_p(x,y)A_p(u-x,v-y),
    \end{aligned}
\end{equation}
where $\boldsymbol{A}_p$ is $\boldsymbol{A}$ zero-padding to doubly the original size.
It could be observed that $\boldsymbol{T}$ is the auto-convolution of $\boldsymbol{A}_p$. Thus, the convolution theorem could be used to get the frequency form of Eq.~(\ref{Temp_def}) as follows:
\begin{equation}
    \boldsymbol{T}=IFFT[FFT(\boldsymbol{A}_p)FFT(\boldsymbol{A}_p)],
\end{equation}
where $FFT$ represents the fast Fourier transform and $IFFT$ is the corresponding inverse transform. Considering $\boldsymbol{T}$ is the doubly upsampling of $\boldsymbol{S}$, the symmetry $\boldsymbol{S}$ of 2D matrix $\boldsymbol{A}$ can be calculated by:
\begin{equation}\label{S_freq}
    \boldsymbol{S}=\mathcal{D}(IFFT[FFT(\boldsymbol{A}_p)FFT(\boldsymbol{A}_p)]),
\end{equation}
where $\mathcal{D}(\cdot)$ is a downsampling function scaling its input matrix to half size. Using the frequency form of auto-convolution function, the computational cost of calculating symmetry $\boldsymbol{S}$ is greatly reduced from Eq.~(\ref{S1_def}) to Eq.~(\ref{S_freq}).

\rev{It should be noted that compared with \cite{loc_distortion}, OAcode proposes a comprehensive restoration strategy for the perspective distortion in case of large information capacity, which is optimized for the 2D barcode scanning process, including the estimation of perspective transformation in the absence of symmetric peaks (Sec.~\ref{PT_restortion}) and the demodulation enhancement scheme for lens distortion. (Sec.~\ref{sec:enhanced_demodulation}).}

\section{OAcode Generation}
Fig.~\ref{fig:generation} is the flowchart of the generation process of OAcode. The input message is converted to the bit sequence to generate the data unit. Then, the data unit is flipped to generate one symmetrical matrix, which would be further cropped according to a set of size parameters to generate the prototype matrix. Finally, we blend the prototype matrix and the background image to obtain the OAcode.

\subsection{Data Unit Generation}\label{DU_gen}
The input message is converted to a bit sequence and reshaped to a 2D bit matrix $\boldsymbol{m}$ of size $L_m \times L_m$. Then, $\boldsymbol{m}$ is spread spectrum encoded using $\boldsymbol{p}$ to generate a data matrix $\boldsymbol{D}$, where $\boldsymbol{p}$ is the spread spectrum module, a bipolar matrix with $\pm 1$ elements alternately arranged like a chessboard. In the spread spectrum process, bit `1'  in the 2D bit matrix $\boldsymbol{m}$ is presented as $+1 \times \boldsymbol{p}$ in the data matrix $\boldsymbol{D}$, and bit `0' in $\boldsymbol{m}$ is represented as $-1 \times \boldsymbol{p}$ in $\boldsymbol{D}$. \rev{According to the spread spectrum theory, transmitting over a larger bandwidth could increase the robustness against external narrowband interference \cite{Theory_SS,CHAPMAN201535}. Thus, we could use a larger spread spectrum module to obtain better robustness against imaging distortions, e.g., blur, noise, and compression, in the screen-camera process.} Finally, a random bipolar matrix is used as the mask matrix $\boldsymbol{K}$ to mask $\boldsymbol{D}$ to generate the data unit $\boldsymbol{U}$:
\begin{equation}
    U(i,j)=D(i,j)K(i,j),
\end{equation}
where $i$ and $j$ denote the index of these matrices. The masking operation makes the distribution of black and white pixels in data unit $\boldsymbol{U}$ more random, which is beneficial to the OAcode extraction. Additionally, the mask matrix $\boldsymbol{K}$ and the spread spectrum module $\boldsymbol{p}$ would be leveraged to estimate the parameters needed in the extraction process, such as the information capacity of OAcode.

\begin{figure*}[t]
  \begin{center}
  \includegraphics[width=\linewidth]{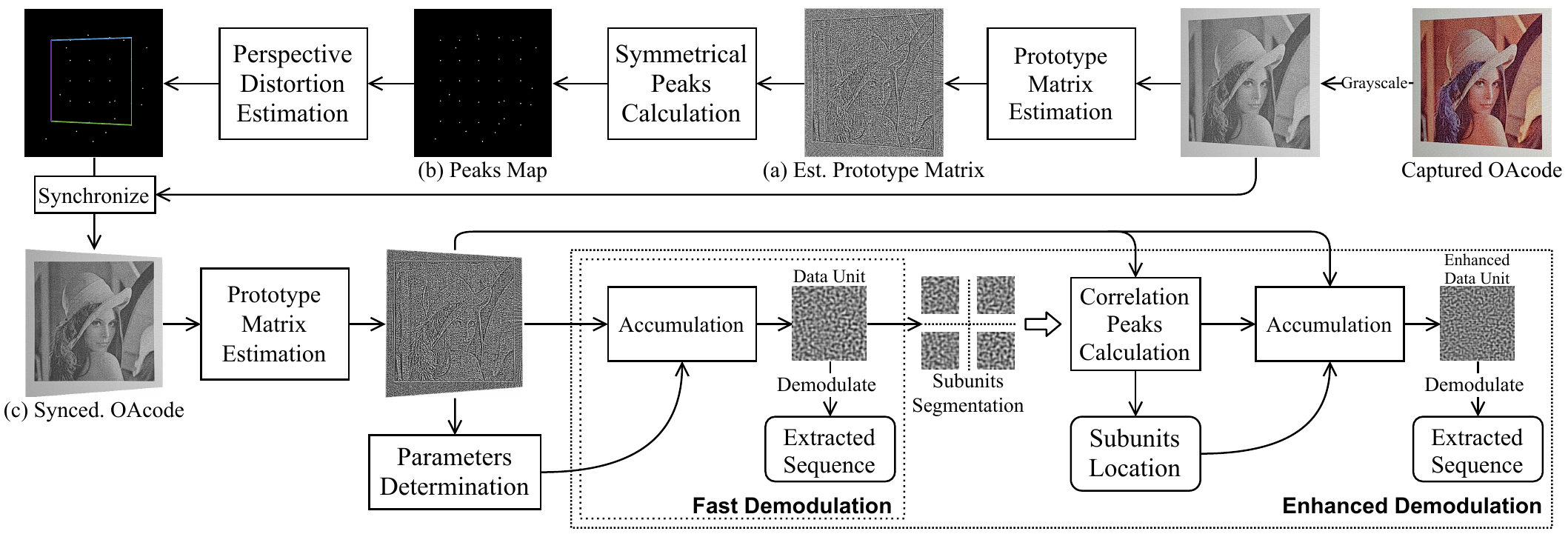}
  \end{center}
\caption{The extraction process of OAcode. The input captured OAcode is used to estimate the prototype matrix in (a), whose symmetrical peaks are calculated to generate the peaks map (b). Then, based on the peaks map, parameters of the perspective distortion could be estimated to synchronize OAcode. Finally, implementing two-stage demodulation on the synchronized OAcode (c) to extract the message sequence.}
\label{fig:extraction}
\end{figure*}

\subsection{Prototype Matrix Generation}\label{proto_gen_section}
The unrotated and unflipped state of the data unit $\boldsymbol{U}$ is defined as the initial state \emph{state 0}. Then, data $\boldsymbol{U}$ with \emph{state 0} is flipped to generate one symmetrical matrix with many symmetrical centers, which are located on the corners of data units. To better illustrate the flipping process, $\boldsymbol{U}$ with \emph{state 0} is represented as a block with a red and bold symbol `d'; see Fig.~\ref{fig:generation}. After the flipping process, that data unit is positioned on the upper left of the global symmetrical center. This positional relationship would be used in the parameter determination strategy of the OAcode extraction process.

Then, the symmetrical matrix is centrally cropped to generate the prototype matrix with the target size, which is determined by two size parameters, including the pixel per inch (PPI) $p$ of the display and the target displayed size $L_d$. Specifically, the cropping size $L_c$ can be calculated by: \rev{$L_c=(L_d \times p)/{C_f}$}, where $C_f=2.54$, is the conversion factor for inches and centimeters. For example, if we need to display an OAcode of size $5 \times 5 cm$ in iPad mini 2 with PPI $p=324$, substituting the above parameters, we can obtain the cropping size $L_c=638$, which means the OAcode should be $638\times 638$ pixels size to achieve the target display effect. 

A noteworthy issue is that the determined size of the prototype matrix would restrict the available information capacity, which is a reasonable result because we cannot display one accessible 2D barcode with a large information capacity in one limited region. In the case of OAcode, a larger information capacity means a data unit with a larger size, resulting in fewer flipped data units in the prototype matrix with the fixed size $L_c$. Yet for better OAcode detection, there should be enough symmetrical centers, i.e., enough data units in OAcode. The size parameters and the information capacity of OAcode restrict each other. In this paper, for better extraction, the parameters we selected make the prototype matrix $\boldsymbol{P}$ contain $5 \times 5$ or more symmetrical centers, i.e., more than $5 \times 5$ data units in $\boldsymbol{P}$.

\subsection{Blending}\label{blending}
Now we have a bipolar prototype matrix $\boldsymbol{P}$, and it is blended with the input background image $\boldsymbol{I}$ to generate the beautified $\boldsymbol{P}$, i.e., the OAcode $\boldsymbol{O}$. Assuming the background image $\boldsymbol{I}$ is an 8-bit grayscale image, whose values range from 0 to 255 and have been resized to the same size as $\boldsymbol{P}$. The proposed OAcode $\boldsymbol{O}$ could be simply generated by:
\begin{equation}\label{simple_em}
    O(i,j)=I(i,j)+\lambda P(i,j),
\end{equation}
where $\lambda$ is the blending intensity. But considering the clipped intensity, OAcode generated by Eq.~(\ref{simple_em}) would be weakened while the background image has brighter or darker contents. To solve the above problem and balance the visual quality and extraction performance, a modulation method with adaptive intensity is proposed:
\begin{equation}\label{adjusted_em}
    \begin{aligned}
    O(i,j)&=I(i,j)+\lambda P(i,j)-2\lambda(I(i,j)/255-0.5)\\
    &=I(i,j)+\lambda\left(P(i,j)+1-\frac{2}{255}I(i,j)\right).
    \end{aligned}
\end{equation}
OAcode generated by Eq.~(\ref{adjusted_em}) is brighter in the darker image content and darker in the brighter content, which is similar to the result of error diffusion dithering. Moreover, Eq.~(\ref{adjusted_em}) could be calculated by matrix operations with a low computational cost. In the case of the colorful background image, the input image is first converted from $RGB$ channels to $YCbCr$ channels, and its $\boldsymbol{Y}$ component would be taken as $\boldsymbol{I}$ to execute the above operations. Finally, the grayscale OAcode is combined with $\boldsymbol{Cb}$, $\boldsymbol{Cr}$ components to convert back to a colorful OAcode.

\section{OAcode Extraction}
Fig.~\ref{fig:extraction} illustrates the extraction process of the proposed OAcode, in which we assume the OAcode captured by the camera  is the only input. To synchronize the captured OAcode, we first estimate the prototype matrix and calculate its symmetrical peaks, which represent the position of the data unit corners and could be used to estimate the geometric distortions, mainly the perspective distortion, occurring on OAcode. After the restoration of perspective distortion, the same as other 2D barcode schemes, we estimate the necessary parameters to demodulate OAcode, including rotation state, information capacity version, and module size. Based on these parameters, OAcode could be demodulated to a bit sequence. Additionally, to resist lens distortion, we propose enhanced demodulation to further improve extraction accuracy.

\subsection{Symmetrical Peaks Calculation}\label{first position detection}
In this process, taking the captured OAcode as the input, we calculate its symmetrical peaks map $\boldsymbol{M}$ for the following OAcode synchronization.

The captured OAcode is first converted to grayscale OAcode $\boldsymbol{O}^\prime$. Then, to decrease the impact of the background image, we separate the prototype matrix $\boldsymbol{P}$ from OAcode $\boldsymbol{O}^\prime$ according to the near-noise appearance of the prototype matrix. Specifically, the prototype matrix $\boldsymbol{P}$ could be estimated by filtering according to \cite{book_wiener,loc_distortion}:

\begin{equation}\label{original_symmetry_cal}
    \widehat{\boldsymbol{P}}=(\boldsymbol{O}^\prime-\boldsymbol{\mu})\frac{\boldsymbol{\sigma}^2_{\boldsymbol{P}}}{\boldsymbol{\sigma}^2_{\boldsymbol{O}^\prime}},
\end{equation}
where $\widehat{\boldsymbol{P}}$ is the estimated prototype matrix (Fig.~\ref{fig:extraction} (a)), $\boldsymbol{\mu}$ is the local mean of $\boldsymbol{O}^\prime$, and $\boldsymbol{\sigma}^2_{\boldsymbol{P}}$ and $\boldsymbol{\sigma}^2_{\boldsymbol{O}^\prime}$ are the local variance of $\boldsymbol{P}$ and $\boldsymbol{O}^\prime$. It should be noted that $\boldsymbol{\sigma}^2_{\boldsymbol{P}}$ is a constant matrix whose values mainly depend on the intensity $\lambda$ in Eq.~(\ref{simple_em}). Thus, Eq.~(\ref{original_symmetry_cal}) could be simplified as:
\begin{equation}
    \widehat{\boldsymbol{P}}=\frac{\boldsymbol{O}^\prime-\boldsymbol{\mu}}{\boldsymbol{\sigma}^2_{\boldsymbol{O}^\prime}}.
\end{equation}
Then, the symmetry matrix $\boldsymbol{S}$ of $\widehat{\boldsymbol{P}}$ can be calculated by the frequency form of the autoconvolution function mentioned in Sec.~\ref{sec:acnf} as follows:
\begin{equation}\label{Sym_cal}
    \boldsymbol{S}=\mathcal{D}(IFFT[FFT(\widehat{\boldsymbol{P}}_p)FFT(\widehat{\boldsymbol{P}}_p)]),
\end{equation}
where $\widehat{\boldsymbol{P}}_p$ is $\widehat{\boldsymbol{P}}$ zero-padding to doubly the origin size and $\mathcal{D}(\cdot)$ is a downsampling function which scales its input matrix to the half size.

Finally, an adaptive symmetrical peaks filter is adopted to generate the symmetrical peaks map as follows:
\begin{equation}\label{Map_cal}
    \boldsymbol{M}=\left\{\begin{array}{ll}{1} & {, \quad if\quad\boldsymbol{S}>\boldsymbol{\mu}_{\boldsymbol{S}}+\beta\boldsymbol{\sigma}^{2}_{\boldsymbol{S}}} \\ {0} & {, \quad otherwise}\end{array}\right.
\end{equation}
where $\boldsymbol{\mu}_{\boldsymbol{S}}$ and $\boldsymbol{\sigma}^{2}_{\boldsymbol{S}}$ respectively denote the local mean and variance of $\boldsymbol{S}$. The coefficient $\beta$ is an adjustable parameter controlling the number of filtered symmetrical peaks. As mentioned in Sec.~\ref{proto_gen_section}, there should be more than $25$ symmetrical centers in the prototype matrix $\boldsymbol{P}$. Thus, as shown in Fig.~\ref{fig:extraction} (b), considering potential noise peaks, we adjust $\beta$ until the peaks map $\boldsymbol{M}$ has slightly more than the expected $25$ peaks.

\subsection{Perspective Distortion Restoration}\label{PT_restortion}
In this process, we estimate perspective distortion parameters from symmetrical peaks map $\boldsymbol{M}$ and apply the corresponding restoration to the captured OAcode.

\begin{figure}[t]
    \begin{center}
    \subfloat[Line detection result of $\boldsymbol{M}$.]{
        {\centering\includegraphics[width=1.6 in]{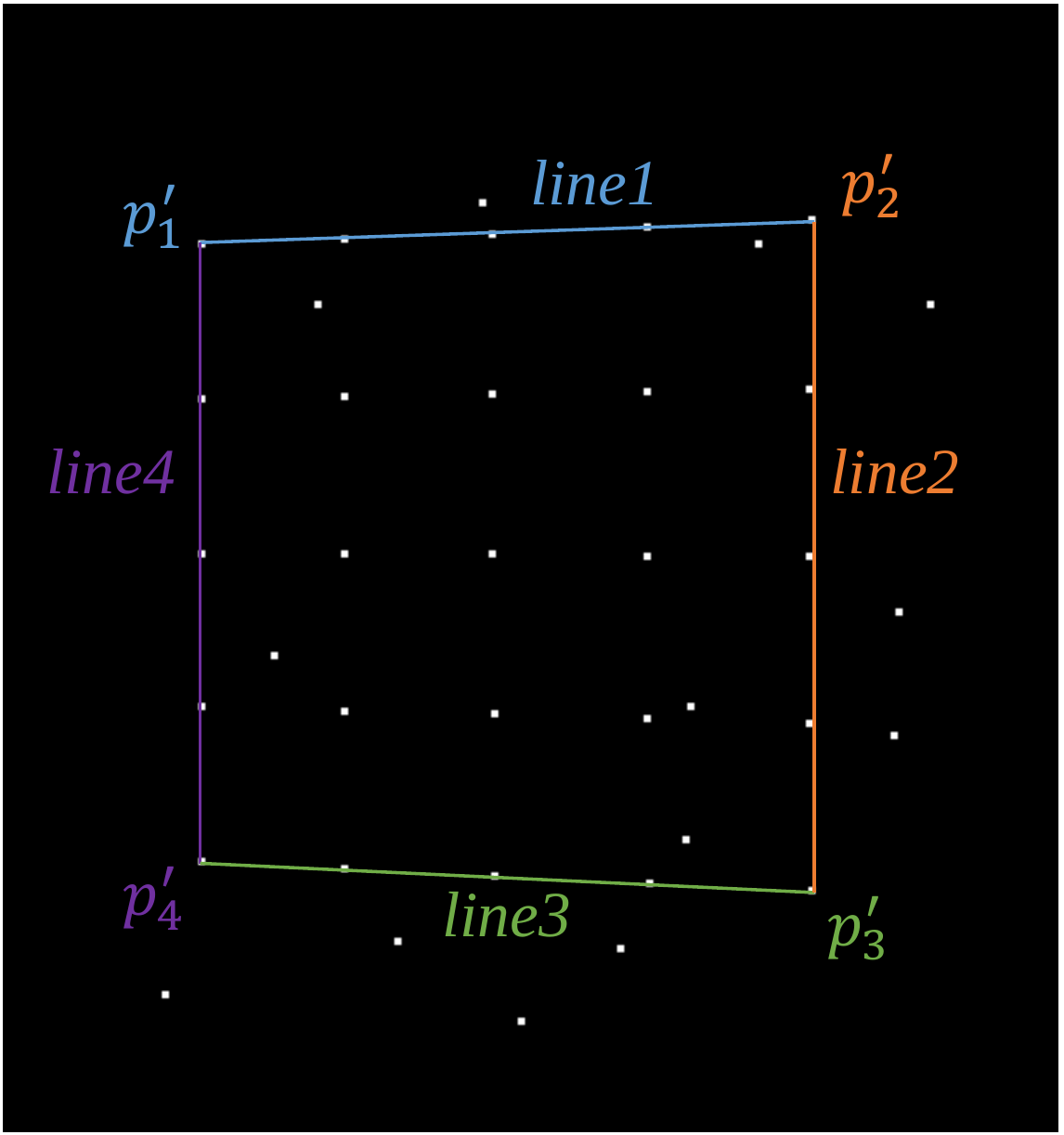}
        }
    }
    \subfloat[Another result of distorted $\boldsymbol{M}$.]{
        {\centering\includegraphics[width=1.6 in]{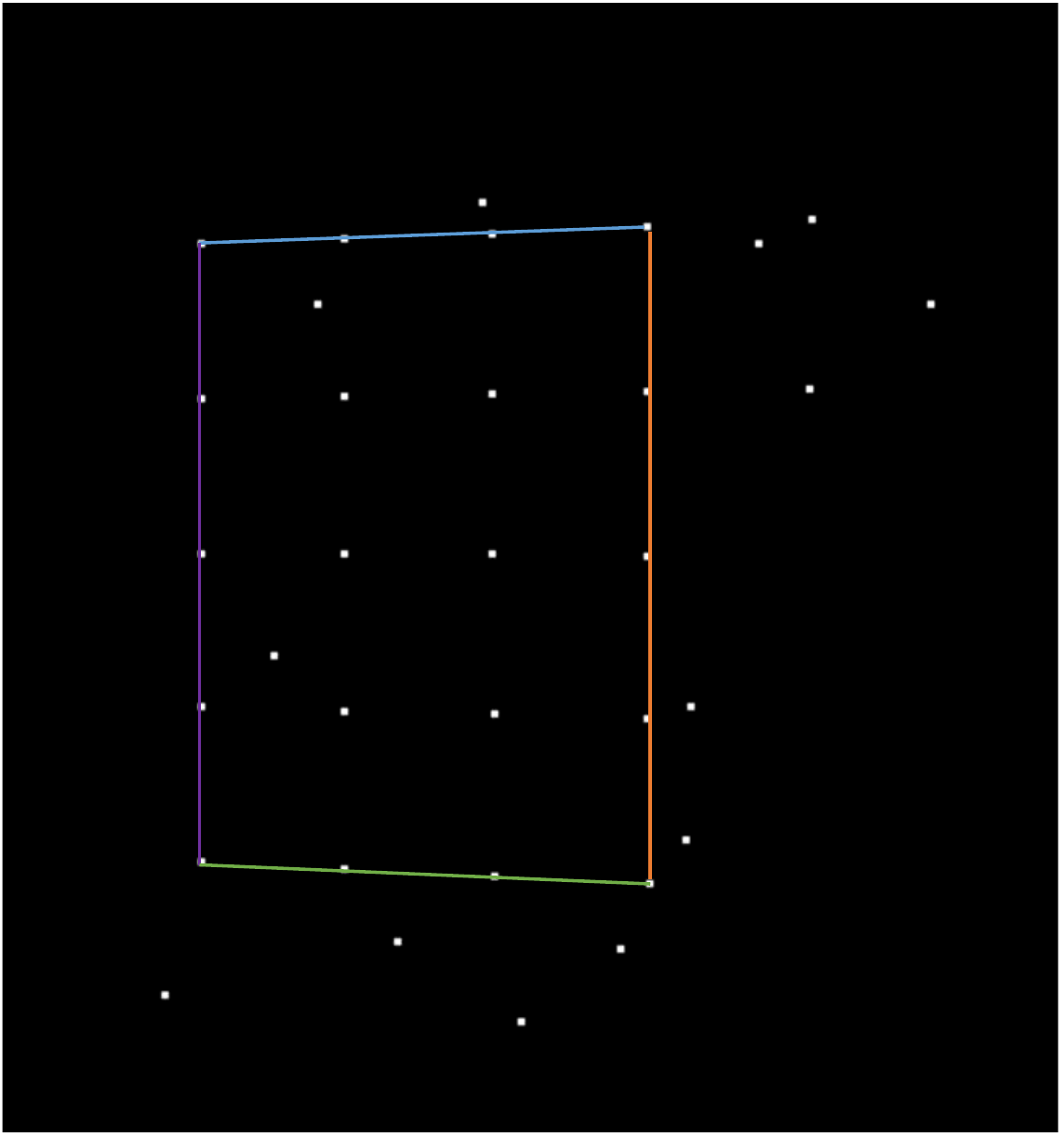}
        }
    }
    \end{center}
\caption{Two different straight line detection results of similar peaks maps $\boldsymbol{M}$s. The peak map in (b) misses some peaks on the right bottom, resulting in the detected right straight line being closer to the center.
}
\label{line_detection}
\end{figure}

For the peaks map $\boldsymbol{M}$, the first technical question is: \emph{How to distinguish symmetrical peaks and noise peaks?} Assuming that the peaks generated by noise are randomly distributed, we can take advantage of the regular distribution of symmetrical peaks, which comes from the grid-type distribution of data units, to distinguish symmetrical peaks. Specifically, these symmetrical peaks represent the corners of data units, so they are also on the edges of data units and are distributed on different straight lines in groups. Thus, we use Hough transform to detect straight lines, which pass through these symmetrical peaks. As shown in Fig.~\ref{line_detection} (a), in all detected lines, we select four outermost lines \emph{line1}, \emph{line2}, \emph{line3}, \emph{line4} and calculate their intersections $p_1^\prime$, $p_2^\prime$, $p_3^\prime$, $p_4^\prime$, which constitute the known points set $\rm P^\prime$. To restore perspective distortion, we also need the target points set $\rm P$, whose points $p_1$, $p_2$, $p_3$, $p_4$ have the undistorted original coordinates of points in $\rm P^\prime$.

In the ideal case, points in known points set $\rm P^\prime$ are the outermost four data unit corners of OAcode, whose relative positions are easily determined. However, considering the serious distortion in the camera-shooting process, sometimes a few peaks are missed in $\boldsymbol{M}$. Fig.~\ref{line_detection} (b) shows another possible line detection result when missing some symmetrical peaks. Thus, the second technical question is: \emph{How to determine the target point set $\rm P$ under different line detection results?} For example, Fig.~\ref{line_detection} (a) and (b) have two different target point set $\rm P$. Our solution takes advantage of the fact that data units are equally spaced, which means the distance between every two symmetrical peaks on the same line is similar even after perspective distortion. We record all distances between every two adjacent points on \emph{line1$\sim$ line4} and use K-means clustering to obtain the biggest distance cluster, whose average value is taken as the estimated side length of the data unit, denoted as $L_u$. Denoting the length of \emph{line1$\sim$ line4} as $L_1$$\sim$$L_4$, the target point set $\rm P$ can be calculated by:
\begin{equation}
    \begin{aligned}
    p_1&(x_1,y_1)=(x_1,y_1)\\
    p_2&(x_2,y_2)=\left(x_1+\mathcal{R}\left(\frac{L_1+L_3}{2L_u}\right)L_u,y_1\right)\\
    p_3&(x_3,y_3)=\left(x_1+\mathcal{R}\left(\frac{L_1+L_3}{2L_u}\right)L_u,y_1+\mathcal{R}\left(\frac{L_2+L_4}{2L_u}\right)L_u\right)\\
    p_4&(x_4,y_4)=\left(x_1,y_1+\mathcal{R}\left(\frac{L_2+L_4}{2L_u}\right)L_u\right),\\
    \end{aligned}
\end{equation}%
where $\mathcal{R}(\cdot)$ is the rounding function. There is no specific constraint to the coordinates of $p_1$, and we make $p_1=p_1^\prime$ empirically. Finally, substituting all coordinates $(x^\prime, y^\prime)$ and $(x, y)$ of the point in $\rm P^\prime$ and $\rm P$ into the following equations:
\begin{equation}\label{pt_parameters}
    \begin{aligned}
    x&=\frac{a_1 x^\prime+b_1 y^\prime + c_1}{a_0 x^\prime + b_0 y^\prime +1}\\
    y&=\frac{a_2 x^\prime+b_2 y^\prime + c_2}{a_0 x^\prime + b_0 y^\prime +1},
    \end{aligned}
\end{equation}
we can solve these equations to obtain all needed parameters $a_0$, $b_0$, $a_1$, $b_1$, $c_1$, $a_2$, $b_2$, $c_2$ for the perspective distortion restoration. As shown in Fig.~\ref{fig:extraction} (c), the captured OAcode would be synchronized to the square state.

\subsection{Parameters Determination}\label{parameters_determination}
To demodulate the restored OAcode, three parameters should be determined, including the rotation state, the information capacity, and the spread spectrum module. In most traditional and aesthetic 2D barcode schemes, these parameters are detected from visible position detection patterns. For the proposed OAcode without position detection patterns, we design a parameter determination strategy based on the statistical characteristics of OAcode.

First, we select a specific data unit in OAcode for analysis. The synchronized OAcode in the previous subsection is used to estimate a synchronized prototype matrix following the same process mentioned in Sec.~\ref{first position detection}, which is denoted as $\widehat{\boldsymbol{P}}$ for notational simplicity. Without the perspective distortion, the symmetry of the synchronized prototype matrix $\widehat{\boldsymbol{P}}$ would be significantly enhanced. Therefore it is reasonable to assume that all the symmetrical peaks are detected by Eq.~(\ref{Sym_cal}) and Eq.~(\ref{Map_cal}), i.e., all data units $\boldsymbol{U}$s in $\widehat{\boldsymbol{P}}$ could be located. Then, the data unit on the upper left of the global symmetrical center \textbf{G} is selected as the data unit to be analyzed, denoted as $\boldsymbol{U}_a$. 

\begin{figure}[t]
    \begin{center}
    \subfloat[\emph{state 0}.]{
        {\centering\includegraphics[width=1.5 in]{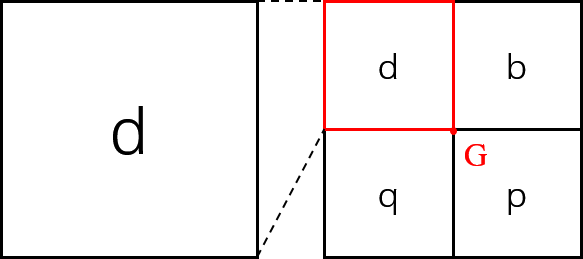}
        }
    }
    \subfloat[\emph{state 1}.]{
        {\centering\includegraphics[width=1.5 in]{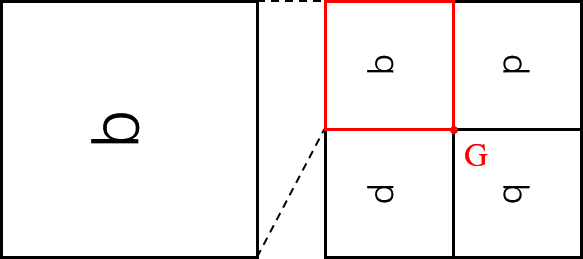}
        }
    }
    \end{center}
\caption{Two possible states of the analyzed data unit $\boldsymbol{U}_a$, i.e., the data unit on the upper left of the global symmetrical center \textbf{G} of the OAcode.
}
\label{states}
\end{figure}

According to the setting in Sec.~\ref{proto_gen_section}, there are two possible states of $\boldsymbol{U}_a$, as shown in Fig.~\ref{states}. When OAcode is not rotated or rotated $180^\circ$, $\boldsymbol{U}_a$ would be the original state, \emph{state 0}. If OAcode is rotated $90^\circ$ or $270^\circ$, $\boldsymbol{U}_a$ would be the \emph{state 1}. The information capacity and size of the spread spectrum matrix $\boldsymbol{p}$ depend on the setting in the OAcode generation process. In this paper, OAcode has two capacity settings, denoted by \emph{capacity 0} and \emph{capacity 1}, and three types of spread spectrum module size, denoted by \emph{module 0}, \emph{module 1}, and \emph{module 2}.
 
For all possible \emph{state i}, \emph{capacity j}, and \emph{module k}, the confidence $C_{ijk}$ that these parameters are correct could be calculated by following steps. According to \emph{state i}, $\boldsymbol{U}_a$ is first rotated and flipped back to \emph{state 0}. Then, $\boldsymbol{U}_a$ is scaled to an expected size determined by \emph{capacity j} and \emph{module k}. For example, in the case of \emph{capacity 0} and \emph{module 0}, the data unit is generated with $L_m \times L_m$ spread spectrum modules $\boldsymbol{p}$s, where $L_m=32$ and $\boldsymbol{p}=\left[ \begin{smallmatrix}
     -1 & \ 1 \\
     \ 1 & -1
\end{smallmatrix} \right]$. Thus, the expected size of $\boldsymbol{U}_a$ is $64 \times 64$. Then, we use the same mask matrix $\boldsymbol{K}$ as Sec.~\ref{DU_gen} to demask $\boldsymbol{U}_a$ to obtain the corresponding data matrix $\boldsymbol{D}_a$:
\begin{equation}
    D_{a}(x,y)=U_{a}(x,y)K(x,y),
\end{equation}
where $x$ and $y$ denote the index of these matrices. If all parameters are correct, data matrix $\boldsymbol{D}_a$ would be a matrix composed of spread spectrum matrix $\boldsymbol{p}$ and $-\boldsymbol{p}$. But if any parameter is wrong, $\boldsymbol{D}_a$ would be the result of a wrong demasking, close to a random matrix. We divide $\boldsymbol{D}_a$ into $L_m \times L_m$ non-overlapping blocks with the same size as $\boldsymbol{p}$. Denoting the block as $\boldsymbol{D}_a^{x,y}$, the confidence $C_{ijk}$ is calculated by:
\begin{equation}
    C_{ijk}=\sum_x^{L_m}\sum_y^{L_m} \left| \langle \overline{\boldsymbol{D}}_a^{x,y} , \boldsymbol{p} \rangle_{\mathbf{F}} \right|,
\end{equation}
where $\overline{\boldsymbol{D}}_a^{x,y}$ is the normalized $\boldsymbol{D}_a^{x,y}$ and $\langle\ \cdot\ ,\ \cdot\ \rangle_{\mathbf{F}}$ is the Frobenius inner product: a component-wise inner product of two matrices. Finally, the \emph{state i}, \emph{capacity j} and \emph{module k} of $\boldsymbol{U}_a$ can be determined by:
\begin{equation}
    i, j, k = \arg \max _{i,j,k}\ C_{ijk}.
\end{equation}
The rest data units have the same \emph{capacity} and \emph{module} with $\boldsymbol{U}_a$, and their \emph{states} can be easily determined according to the relative position of themselves and $\boldsymbol{U}_a$.

\subsection{Demodulation}
We propose a two stages demodulation method to meet different requirements of accuracy and computational cost in the demodulation process of OAcode. The first stage conducts a fast demodulation method with less computational cost. In most cases, its demodulation result is accurate enough. When users pursue higher demodulation accuracy, enhanced demodulation in the second stage would be adopted, which considerably decreases demodulation error caused by lens distortion. 

\subsubsection{Fast Demodulation}\label{standard_demodulation}
According to the location and parameters of data units determined in Sec.~\ref{parameters_determination}, we restore all data units back to \emph{state 0} and accumulate them to generate one accumulated data unit $\widehat{\boldsymbol{U}}$. The accumulation operation can efficiently reduce the impact of random noise. Similar to Sec.~\ref{parameters_determination}, we resize the accumulated data unit $\widehat{\boldsymbol{U}}$ to the expected size according to its \emph{capacity} and \emph{module}. Then, the accumulated data unit $\widehat{\boldsymbol{U}}$ is demasked with the mask matrix $\boldsymbol{K}$ to obtain the data matrix $\widehat{\boldsymbol{D}}$, which is divided into $L_m \times L_m$ non-overlapping and normalized blocks $\widehat{\boldsymbol{D}}^{x,y}$. To despread data matrix $\widehat{\boldsymbol{D}}$, the correlation $\rho_{x,y}$ between $\widehat{\boldsymbol{D}}^{x,y}$ and $\boldsymbol{p}$ is calculated by:
\begin{equation}
    \rho_{x,y}=\langle \widehat{\boldsymbol{D}}^{x,y} , \boldsymbol{p} \rangle_{\mathbf{F}},
\end{equation}
where $\langle\ \cdot\ ,\ \cdot\ \rangle_{\mathbf{F}}$ is the Frobenius inner product, $x$ and $y$ represent the block index in $\widehat{\boldsymbol{D}}$, and the spread spectrum module $\boldsymbol{p}$ is determined by the \emph{module} parameter. Then, the message matrix $\boldsymbol{m}^\prime$ can be extracted by:
\begin{equation}
    m^\prime_{x,y}=\left\{\begin{array}{ll}{0} & {, \quad \rho_{x,y}<0} \\ {1} & {, \quad \rho_{x,y} \geq 0}\end{array}\right.
\end{equation}
Finally, we reshape the message matrix to a bit sequence, i.e., the extracted message.

\subsubsection{Enhanced Demodulation}
\label{sec:enhanced_demodulation}
Considering the imperceptible lens distortion in the captured OAcode, the accuracy of the fast demodulation could be further improved by the enhanced demodulation, which reduces the impact of lens distortions. We first discuss how lens distortions impact the demodulation accuracy and then illustrate the implementation of the enhanced demodulation method.

In fast demodulation, all data units in OAcode are accumulated to obtain one accumulated data unit $\widehat{\boldsymbol{U}}$ to enhance the meaningful signal and weaken random noise. However, the lens distortion in different data units is different, which causes misalignments in the data unit accumulation and weakens the accumulation result. Ignoring other factors, the smaller the size of one block in the camera-shooting OAcode, the weaker the lens distortion, and the less the bit error rate caused by misalignments. Inspired by this rule, we design the following enhanced demodulation method.

\begin{figure}[t]
  \begin{center}
  \includegraphics[width=3.4in]{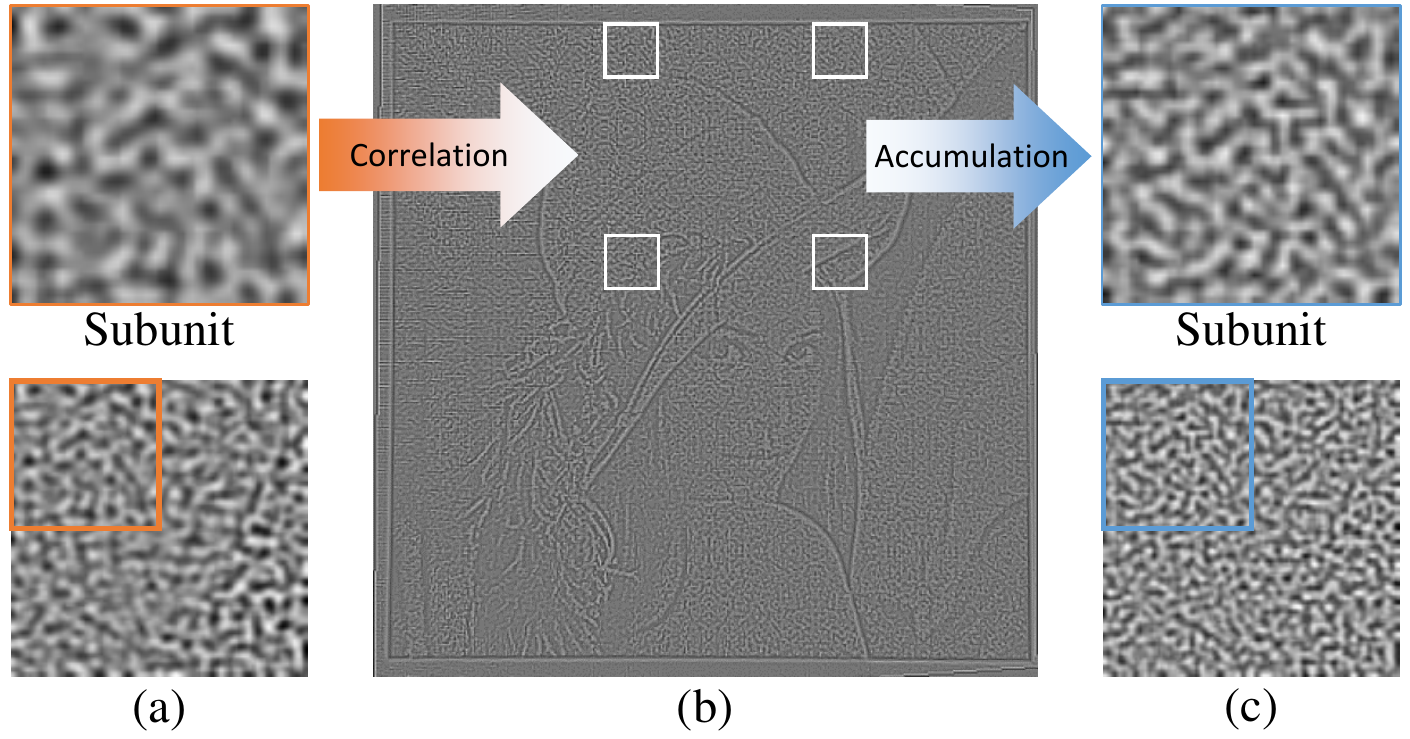}
  \caption{An example of the second position detection of data units. (a) is the accumulated data unit, which is divided into $4$ subunits. (b) is the synchronized prototype matrix, in which white borders are the detection results of one subunit. (c) is the enhanced data unit, generated by the accumulation of these detected subunits.
  }\label{second_detection}
  \end{center}
\end{figure}

In this enhanced demodulation, the steps up to the generation of the accumulated data unit $\widehat{\boldsymbol{U}}$ are the same as the fast demodulation method. Then, $\widehat{\boldsymbol{U}}$ is used for the second position detection of data units rather than direct demodulation. The first position detection of data units is in Sec.~\ref{first position detection}, where the corners of data units are detected as the symmetrical peaks by the auto-convolution function. The second position detection of data units is based on the correlation peaks calculated by the cross-correlation between the accumulated data unit $\widehat{\boldsymbol{U}}$ and the synchronized prototype matrix $\widehat{\boldsymbol{P}}$. To decrease the impact of lens distortions, we take the smaller subunit to execute the second position detection. Specifically, as shown in Fig.~\ref{second_detection} (a), we divide the accumulated data unit $\widehat{\boldsymbol{U}}$ into $n \times n$ non-overlapping subunits $\widehat{\boldsymbol{U}}^{x,y}$. Denoting $L_x$ and $W_y$ as the length and the width of the subunit $\widehat{\boldsymbol{U}}^{x,y}$, and $L_U$ as the size of the accumulated data unit $\widehat{\boldsymbol{U}}$, $L_x$ and $W_y$ can be determined by:
\begin{equation}
\begin{aligned}
    L_x&=\left\{\begin{array}{ll}{\lfloor \frac{L_u}{n} \rfloor} & {, \quad x\in [1,n-1]} \\ {L_U-(n-1)\lfloor \frac{L_u}{n} \rfloor} & {, \quad x=n}\end{array}\right.\\
    W_y&=\left\{\begin{array}{ll}{\lfloor \frac{L_u}{n} \rfloor} & {, \quad y\in [1,n-1]} \\ {L_U-(n-1)\lfloor \frac{L_u}{n} \rfloor} & {, \quad y=n}\end{array}\right.
\end{aligned}
\end{equation}
Then the cross-correlation $\boldsymbol{C}_{xy}$ between subunit $\widehat{\boldsymbol{U}}^{x,y}$ and the synchronized prototype matrix $\widehat{\boldsymbol{P}}$ can be calculated by:
\begin{equation}
\boldsymbol{C}_{xy}=IFFT[FFT(\widehat{\boldsymbol{U}}^{x,y})FFT(\widehat{\boldsymbol{P}})^*]
\end{equation}
where $*$ represents the conjugate of the matrix. As shown in Fig.~\ref{second_detection} (b), according to correlation peaks in $\boldsymbol{C}_{xy}$, we could locate the corresponding subunits and accumulate them to generate an enhanced subunit. Repeating the above operations for all possible $x$ and $y$, an enhanced data unit with less lens distortion can be obtained. As shown in Fig.~\ref{second_detection} (c) and (a), the enhanced data unit is much clear and has a much lower bit error rate (1.43\%) than the original one (11.18\%). In the practice, the enhanced subunit is accumulated by only four subunits with the largest four correlation peaks, because the weak correlation peak means an unclear subunit, which brings more interference than improvement. Finally, the enhanced data unit is demodulated by the same process as the one in Sec.~\ref{standard_demodulation}.

\section{Experiment Evaluation}
\subsection{Experimental Methodology}\label{Setting}
In this subsection, we illustrate the default experimental setting. For a specific experiment, except for some rewritten parameters, most experimental parameters are the same as the default setting. 

In the OAcode generation process, the prototype matrix is set to $32 \times 32$ (1024) bits information capacity and modulated by $3 \times 3$ size spread spectrum module $\boldsymbol{p}$. Then, the prototype matrix is cropped to $640 \times 640$ pixels and blended with the background image to generate the tested OAcode, in which the blending intensity $\lambda$ is set to $20$.

In the extraction process, two screens are used to display OAcodes. One is iPad mini 2, whose resolution is $2048 \times 1536$ pixels and PPI (pixel per inch) is 324. And the other is Huawei P30 Pro, having a $2340 \times 1080$ pixels resolution and 398 PPI. For the OAcode with the default size setting of $640 \times 640$ pixels, it would be displayed with a size of $5 \times 5cm$ on iPad min 2 and a size of $4 \times 4cm$ on Huawei P30 Pro. The illuminance of displays is set to $80 \pm 10 lux$. Then, OAcode is captured by two mobile phones, iPhone 11 and Xiaomi Mi 6 with a limited capturing resolution of 720p ($1280 \times 720$ pixels) at the screen-camera distance of $10cm$. The camera-shooting process is done indoors with an illuminance of around $300 lux$.In the enhanced demodulation, if not specified, the accumulated data unit is divided into $2 \times 2$ subunits for the second position detection.

It should be noted that the following performance of OAcode is evaluated with clearly defined experimental parameters so that for new devices or new parameters, some performance of OAcode on them could be reasonably inferred, i.g., the available distance.

\begin{figure}[t]
  \begin{center}
  \includegraphics[width=3.4in]{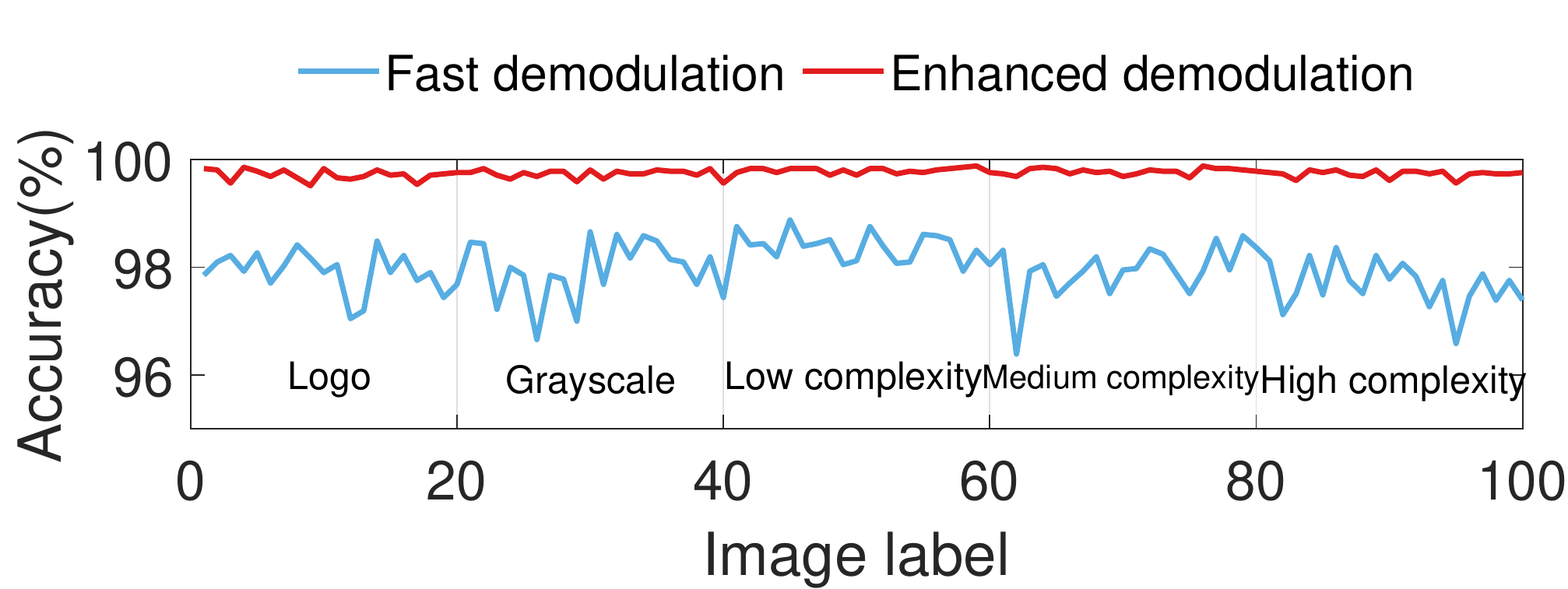}
  \caption{Overall performance of OAcode.}
  \label{overall}
  \end{center}
\end{figure}

\begin{figure}[t]
\setlength{\belowcaptionskip}{0pt}
\setlength{\abovecaptionskip}{0pt}
    \begin{center}
    \subfloat[Label 4]{
        {\centering\includegraphics[width=.54in]{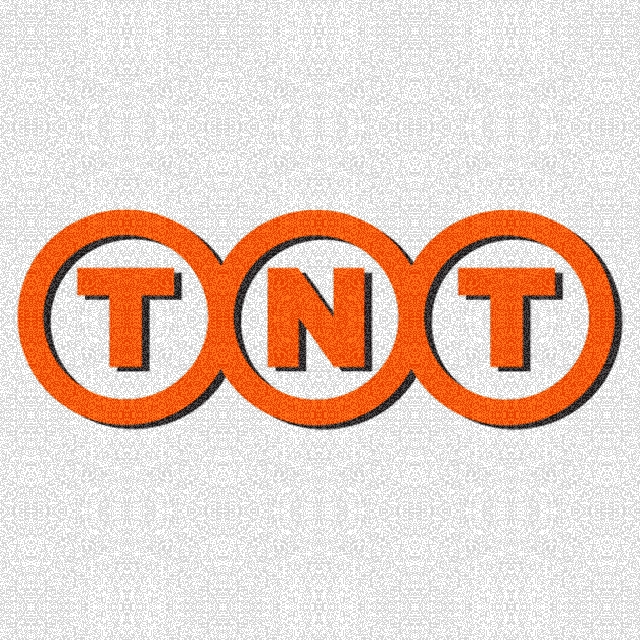}
        }
    }
    \subfloat[Label 22]{
        {\centering\includegraphics[width=.54in]{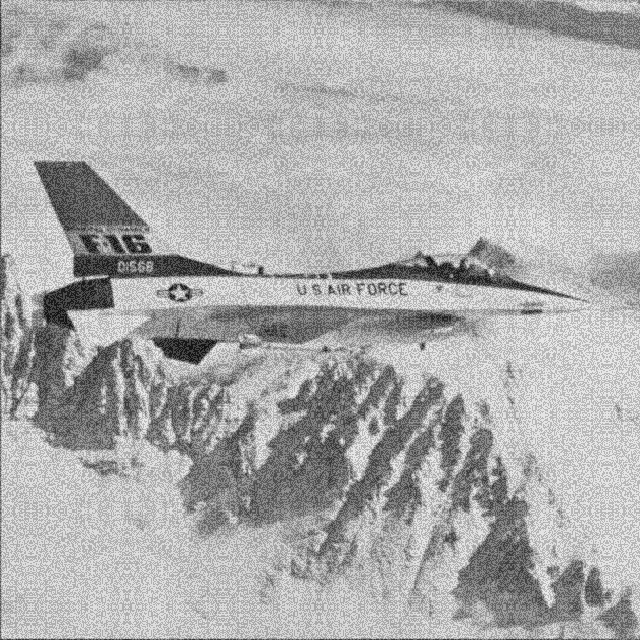}
        }
    }
    \subfloat[Label 59]{
        {\centering\includegraphics[width=.54in]{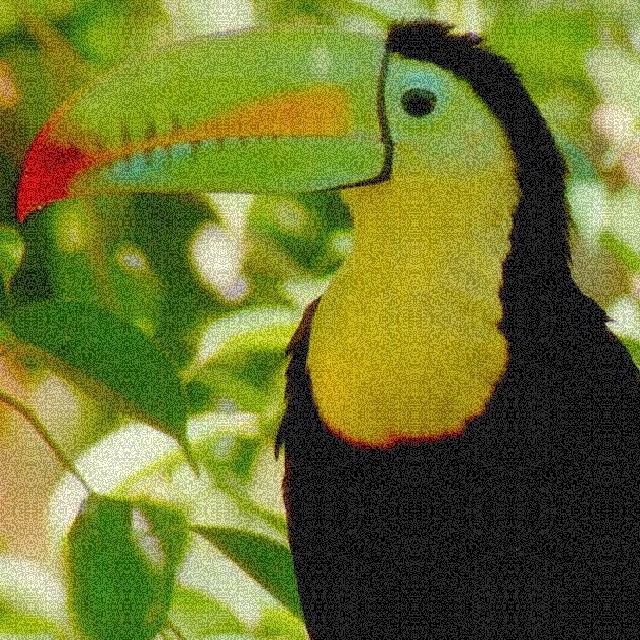}
        }
    }
    \subfloat[Label 76]{
        {\centering\includegraphics[width=.54in]{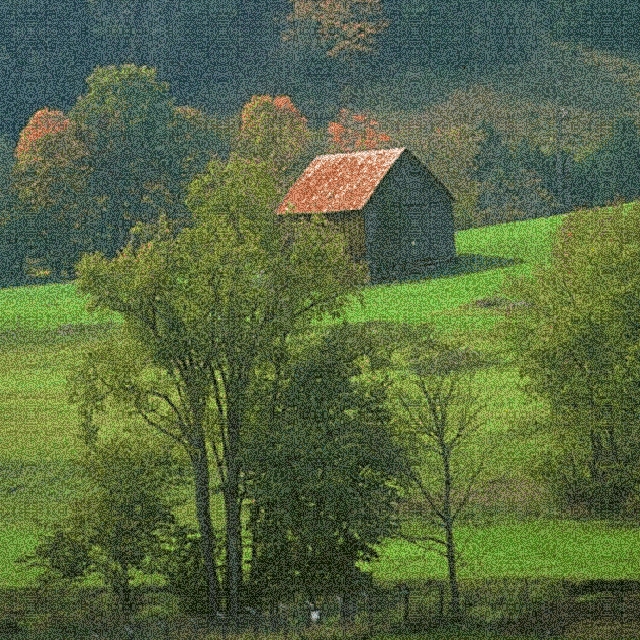}
        }
    }
    \subfloat[Label 86]{
        {\centering\includegraphics[width=.54in]{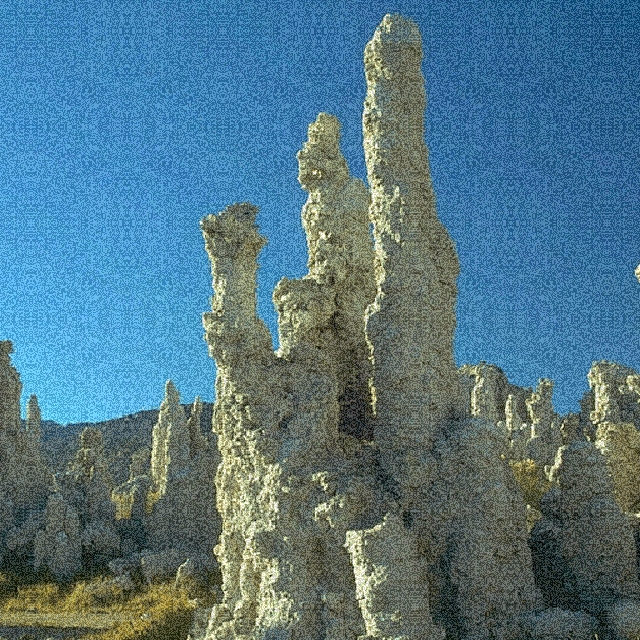}
        }
    }
    \\
    \subfloat[Label 9]{
        {\centering\includegraphics[width=.54in]{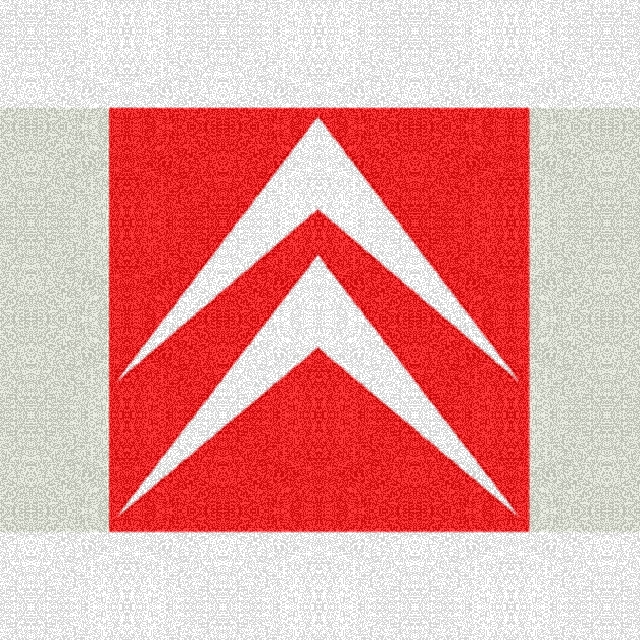}
        }
    }
    \subfloat[Label 40]{
        {\centering\includegraphics[width=.54in]{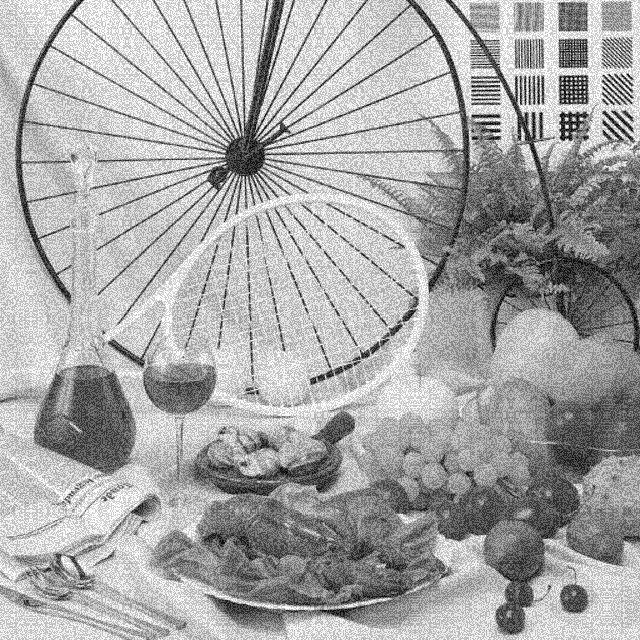}
        }
    }
    \subfloat[Label 50]{
        {\centering\includegraphics[width=.54in]{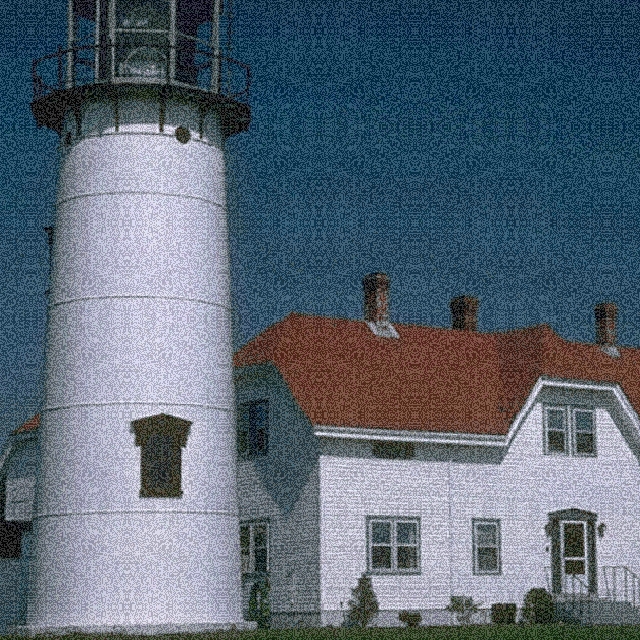}
        }
    }
    \subfloat[Label 75]{
        {\centering\includegraphics[width=.54in]{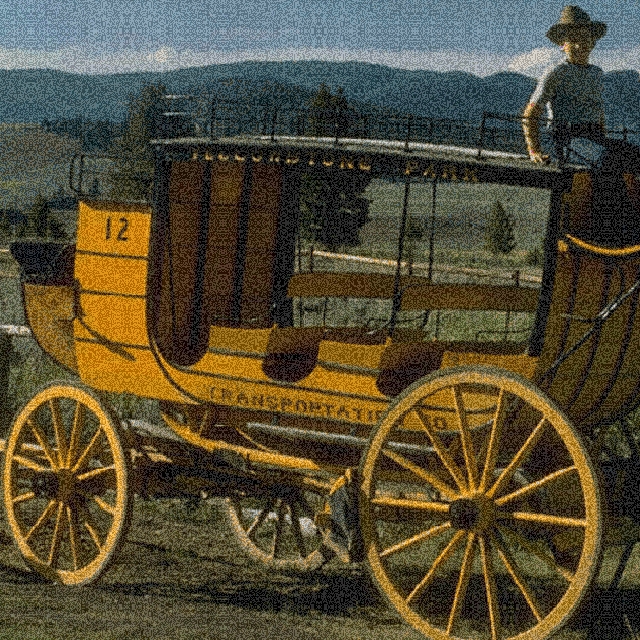}
        }
    }
    \subfloat[Label 95]{
        {\centering\includegraphics[width=.54in]{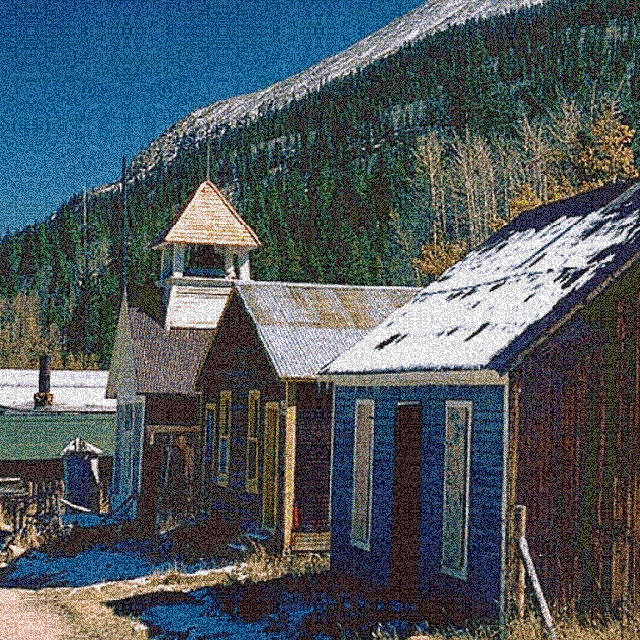}
        }
    }
    \end{center}
    
    \caption{Top row: OAcodes with the highest accuracy in five image categories. Bottom row: OAcodes with the lowest accuracy. 
    }
    \label{samples}
\end{figure}

\subsection{OAcode Performance}
\subsubsection{Overall Performance}
\label{sec:overall_performance}
We select $100$ background images to evaluate the overall performance of the proposed OAcode. To cover most scenarios, the selected background images include $5$ categories, logo, grayscale, low complexity color images, medium complexity color images, and high complexity color images. Each category has $20$ images. The logo images are selected from BelgaLogos Dataset\cite{BelgaLogos}, and the other images come from the CVG-UGR Image database \cite{CVG}. In this paper, we use the multiscale structural similarity index (MS-SSIM) and the peak signal-to-noise ratio (PSNR) to measure the objective visual quality of OAcode. In this subsection, these OAcode have an average MS-SSIM of $0.821$ and an average PSNR of $22.45$. \rev{Additionally, we invite 20 participants to evaluate the visual quality of OAcode with the criteria that the score of the black-and-white QR code and the original background image is $1$ and $10$ respectively. The proposed OAcode obtains a mean opinion score of $8.15$, which is beneficial from its noise-like data area without visible position detection patterns. }

The extraction results are shown in Fig.~\ref{overall}, in which the demodulation accuracy of one OAcode is the average accuracy extracted under four possible screen-camera combinations. The blue line represents the accuracy calculated by fast demodulation, whose minimum is over $96\%$. The red line is the demodulation result of the enhanced demodulation with the minimum exceeding 99.5\%. These extraction results illustrate that the proposed OAcode has a very high extraction accuracy on various background images of different categories.

The images with the lowest and the highest accuracy in each category are selected as representative background images. As shown in Fig.~\ref{samples}, OAcodes generated by these $10$ representative background images are evaluated in the subsequent experiments.

\subsubsection{Detection and Demodulation Range}\label{extraction_range}
\begin{figure}[t]
\setlength{\belowcaptionskip}{0pt}
\setlength{\abovecaptionskip}{0pt}
    \begin{center}
    \subfloat[Impact of distance on the\\detection rate.]{
        {\centering\includegraphics[width=1.65 in]{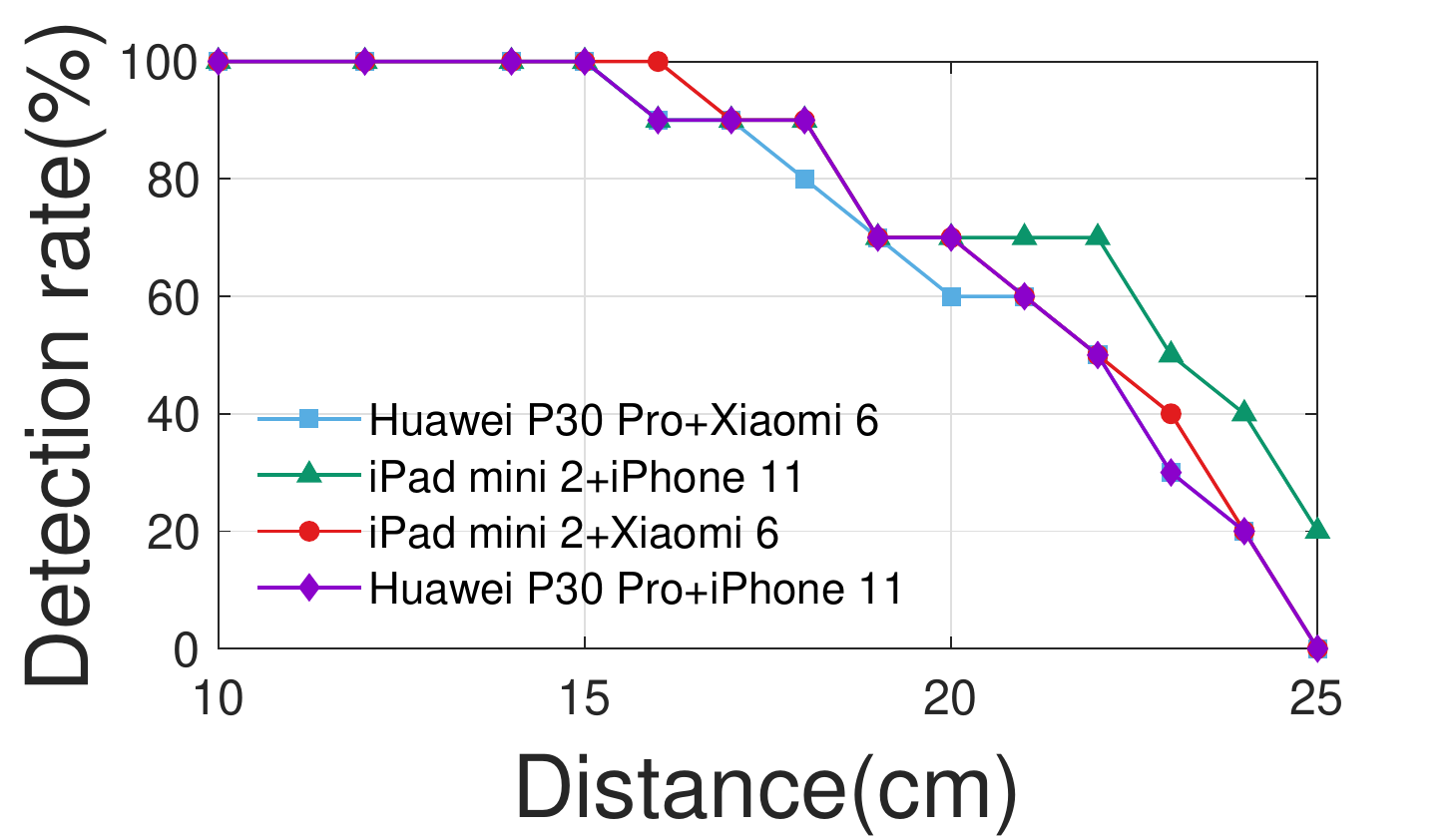}
        }
    }
    \subfloat[Impact of distance on the demodulation accuracy.]{
        {\centering\includegraphics[width=1.55 in]{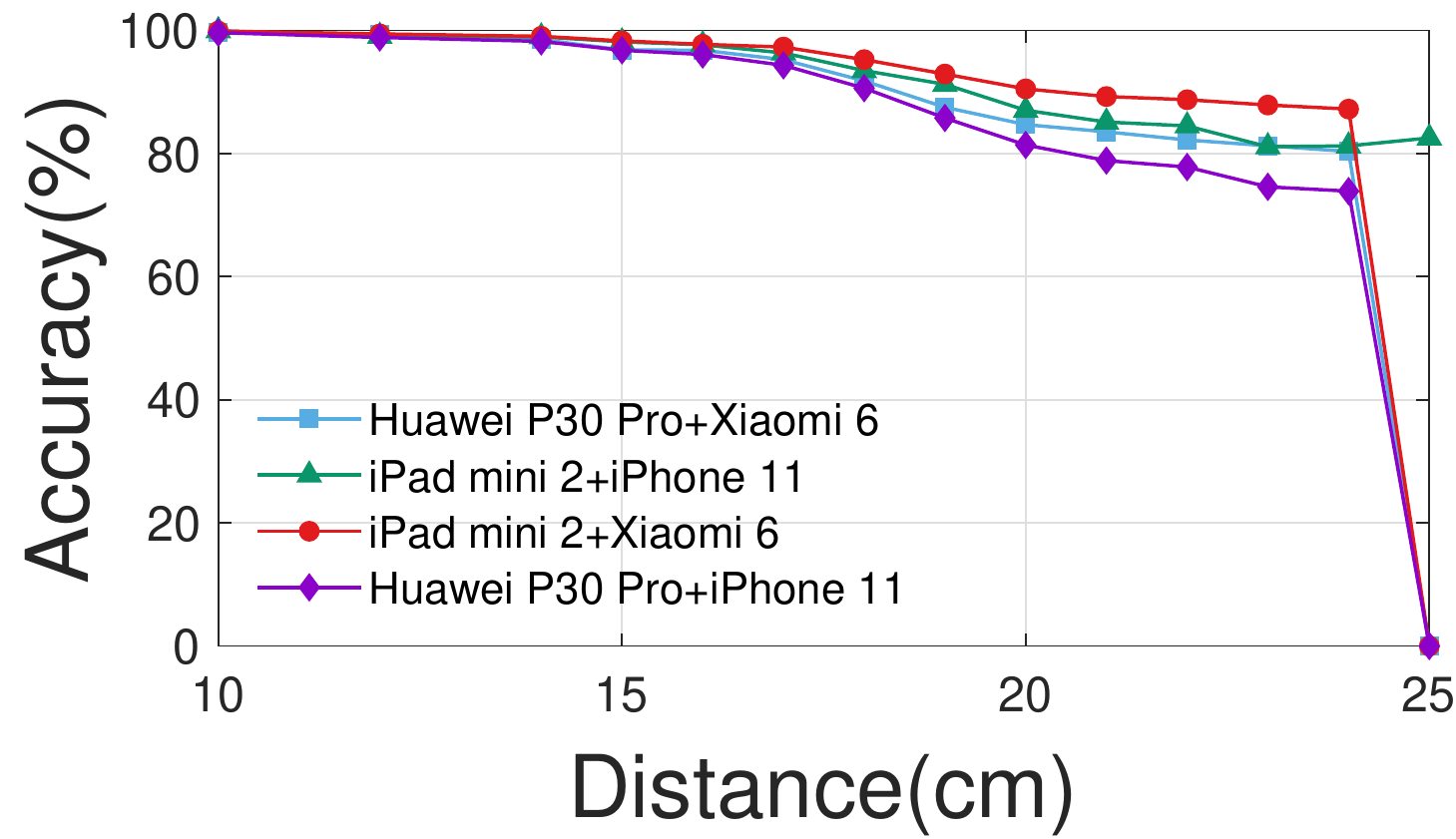}
        }
    }
    \\
    \subfloat[Impact of angle on the detection rate.]{
        {\centering\includegraphics[width=1.6 in]{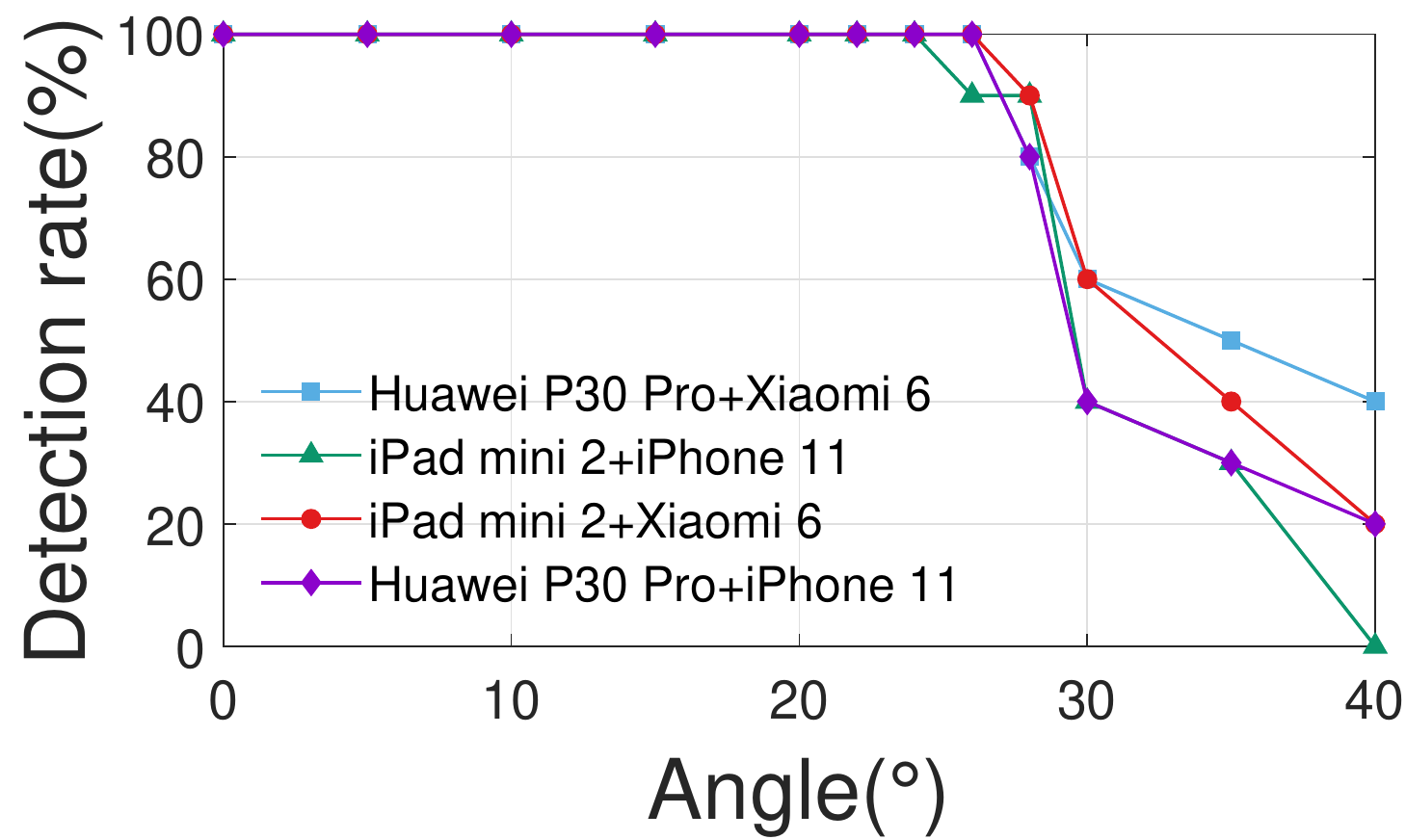}
        }
    }
    \subfloat[Impact of angle on the demodulation accuracy.]{
        {\centering\includegraphics[width=1.6 in]{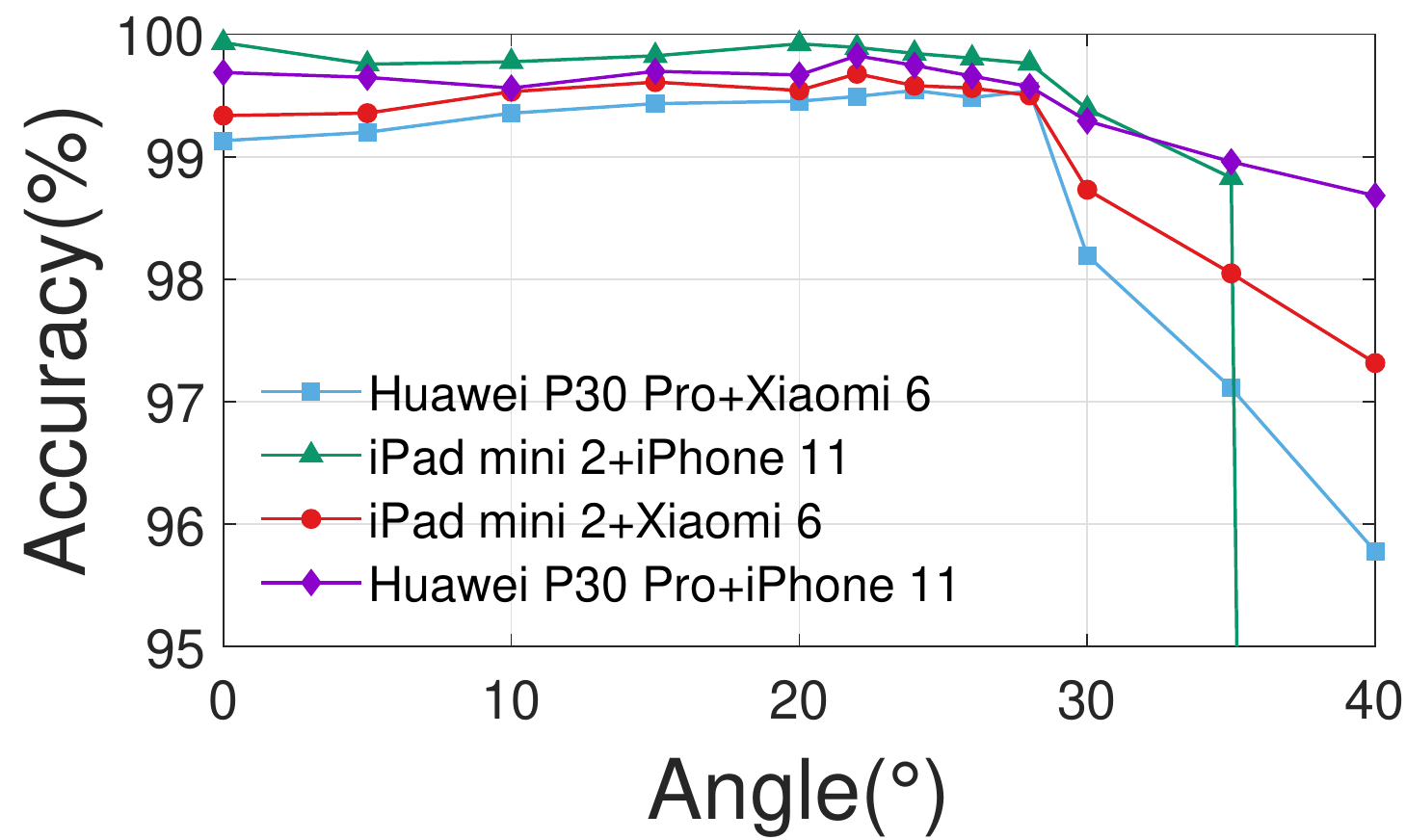}
        }
    }
    \end{center}
\caption{The detection and demodulation range of OAcode.
}
\label{disang}
\end{figure}

To evaluate the performance of OAcode under different camera-shooting conditions, $10$ OAcodes in Fig.~\ref{samples} are captured at various screen-camera distances and angles. The extraction results are shown in Fig.~\ref{disang}, in which (a) and (c) illustrate the detection rate of the captured OAcodes as well as (b) and (d) illustrate the demodulation accuracy. We take the detection rate to represent the probability that the captured OAcode could be resynchronized. Then, the synchronized OAcode is demodulated to obtain the extracted message, whose bit recovery rate is used as the demodulation accuracy. 

According to Fig.~\ref{disang} (a) and (b), at $15cm$ screen-camera distance, OAcode has a $100\%$ detection rate and up to $96.76\%$ demodulation accuracy. Although the detection rate of OAcode drops to $60\%$ at a screen-camera distance of $20cm$, its demodulation accuracy still exceeds $80\%$. Considering using several frames to extract messages is common in 2D barcode applications, the detection rate of OAcode could be theoretically improved to $93.6\%$ using $3$ frames at a distance of $20cm$. Because of the high demodulation accuracy of OAcode, OAcode could be extracted at a distance of $24cm$ by sufficiently scanning. In the angle experiment, the screen-camera distance is fixed as $10cm$. As shown in Fig.~\ref{disang} (c) and (d), at a $25^\circ$ screen-camera angle, OAcode could be detected and demodulated almost without error. Similarly, at a scanning angle of $35^\circ$, OAcode can be theoretically extracted with over $97\%$ demodulation accuracy.

\subsubsection{Impact of Spread Spectrum Module Size}
The size of the spread spectrum module $\boldsymbol{p}$, $\boldsymbol{p}\_size$, has a considerable influence on the performance of OAcode. According to the interconstraint relationship between parameters mentioned in Sec.~\ref{proto_gen_section}, when OAcode is set to $640 \times 640$ pixels and $1024$ bits information capacity, the size of $\boldsymbol{p}$ can only be $2\times 2$, $3\times 3$, or $4 \times 4$ pixels, corresponding to three types of OAcode. Then, based on $10$ representative background images, we generate $3 \times 10$ OAcodes and evaluate their performance differences caused by $\boldsymbol{p}\_size$.

\begin{figure}[t]
    \begin{center}
    \subfloat[Impact of distance on the detection rate.]{
        {\centering\includegraphics[width=1.6 in]{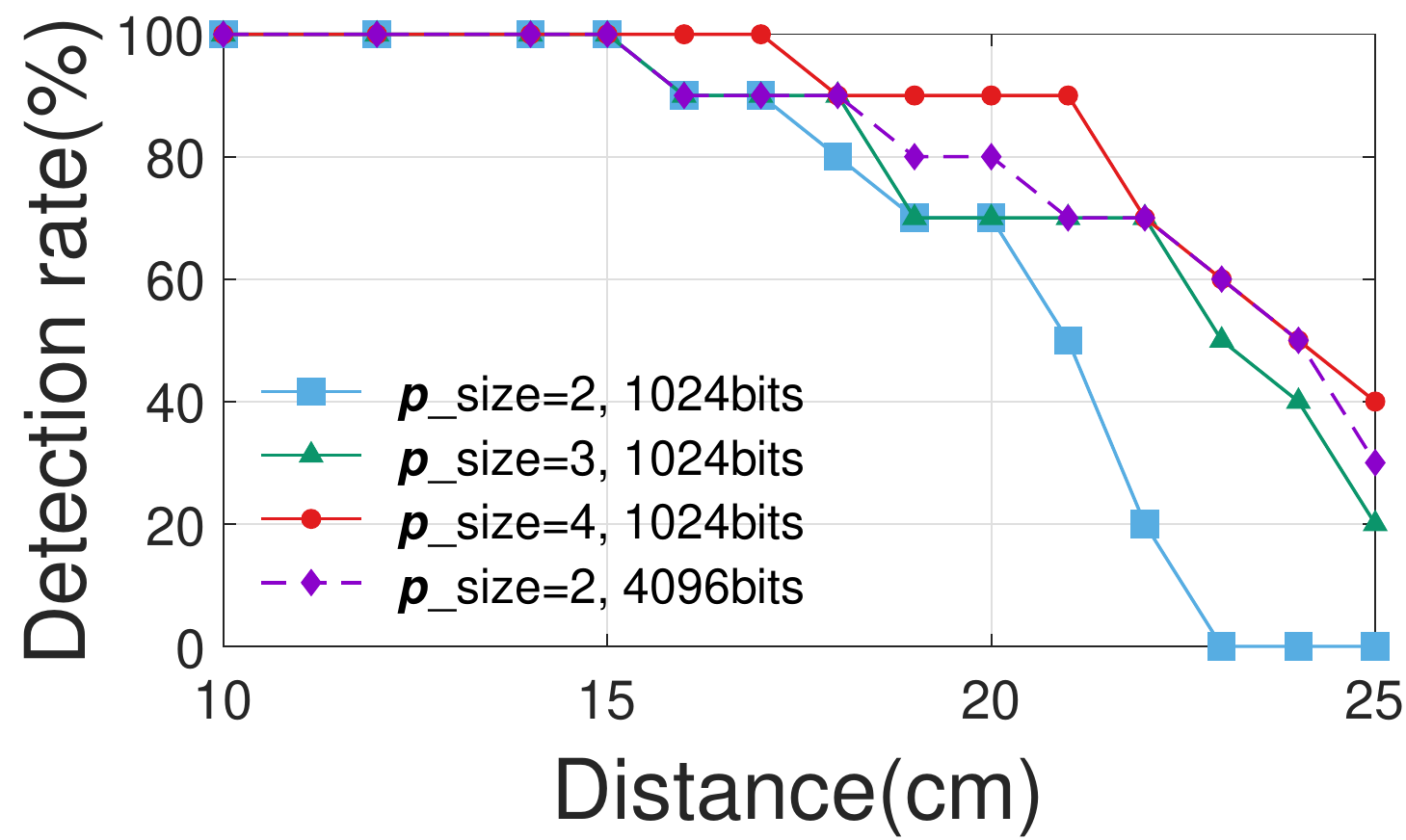}
        }
    }
    \subfloat[Impact of distance on the demodulation accuracy.]{
        {\centering\includegraphics[width=1.6 in]{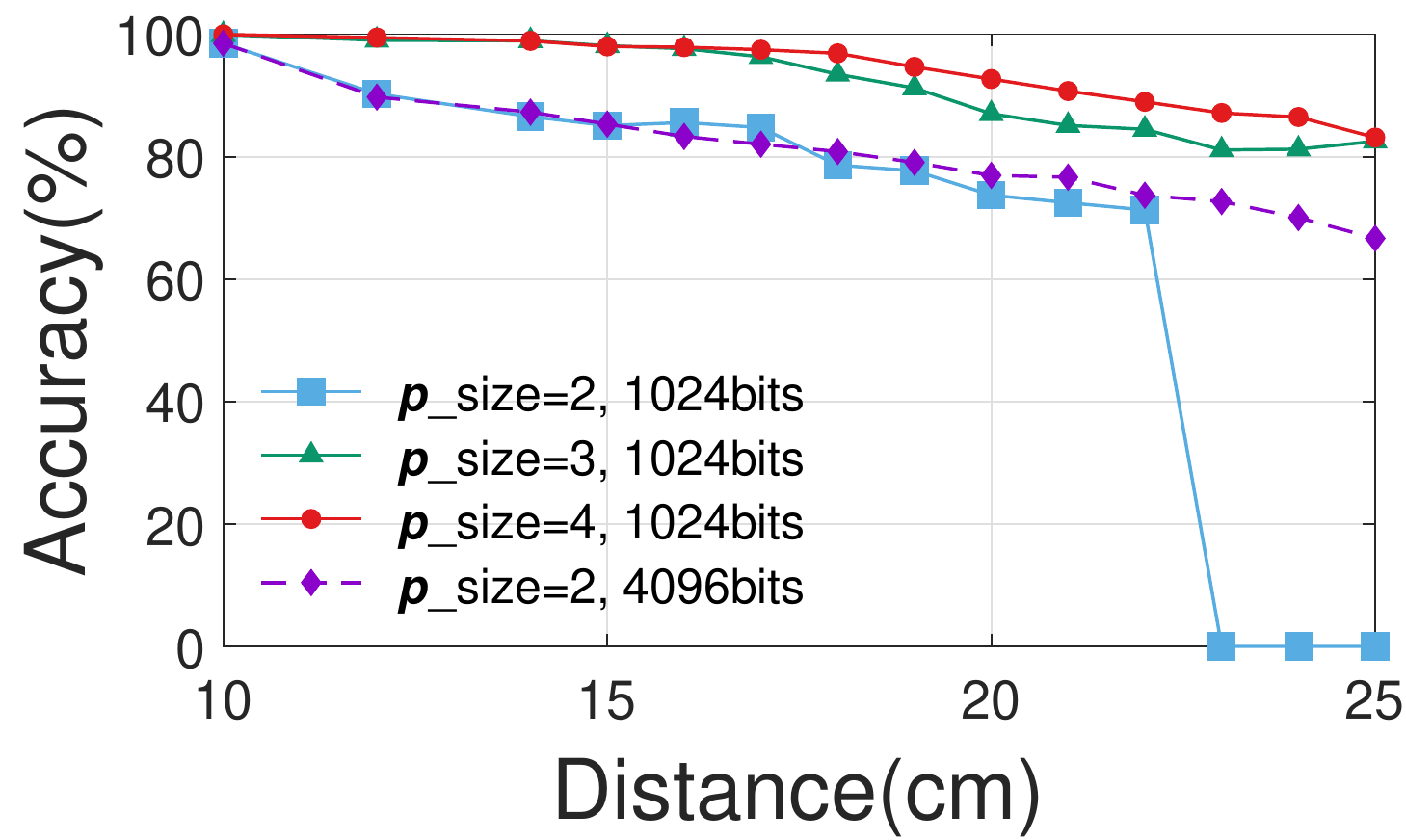}
        }
    }
    \\
    \subfloat[Impact of angle on the detection rate.]{
        {\centering\includegraphics[width=1.6 in]{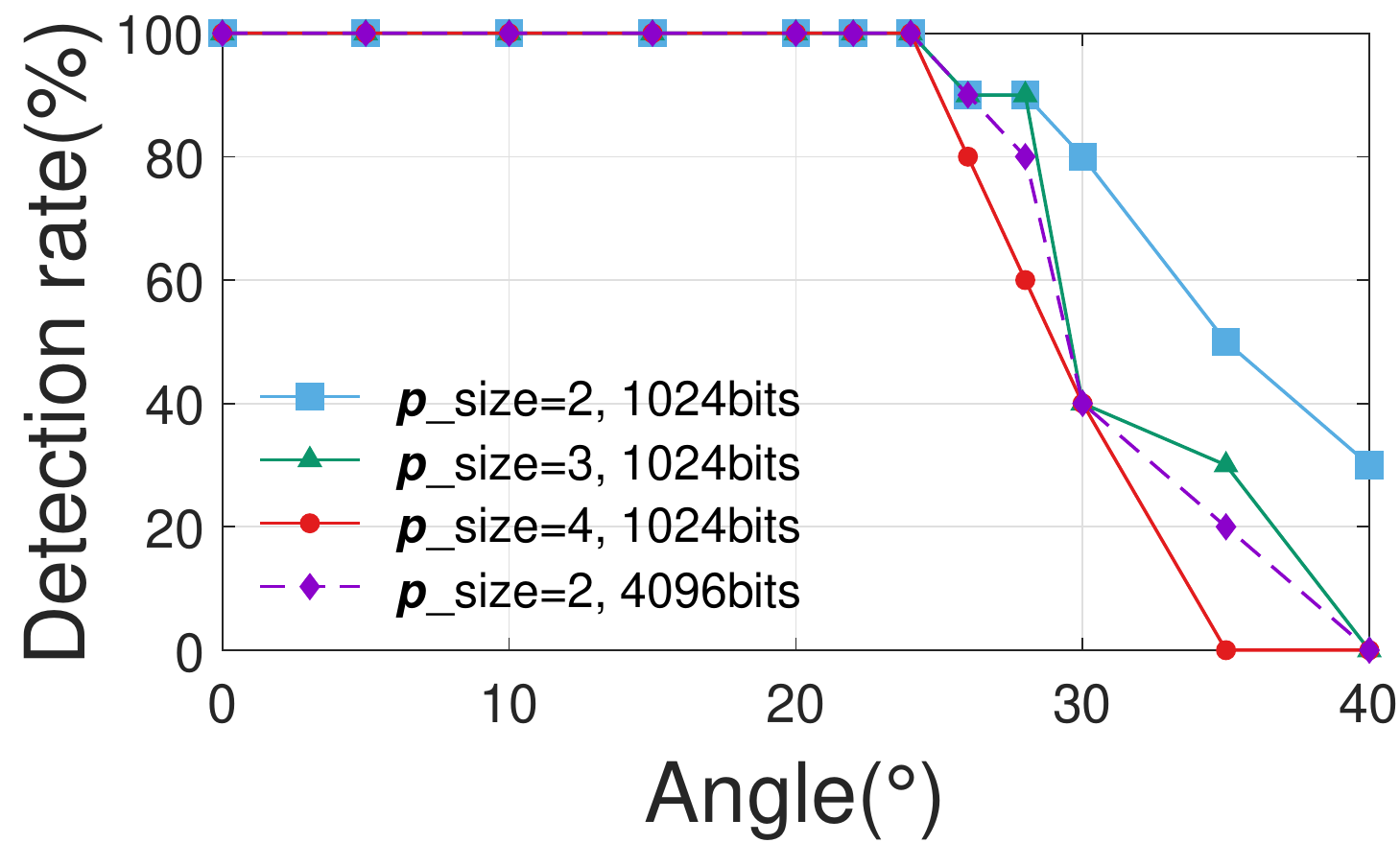}
        }
    }
    \subfloat[Impact of angle on the demodulation accuracy.]{
        {\centering\includegraphics[width=1.6 in]{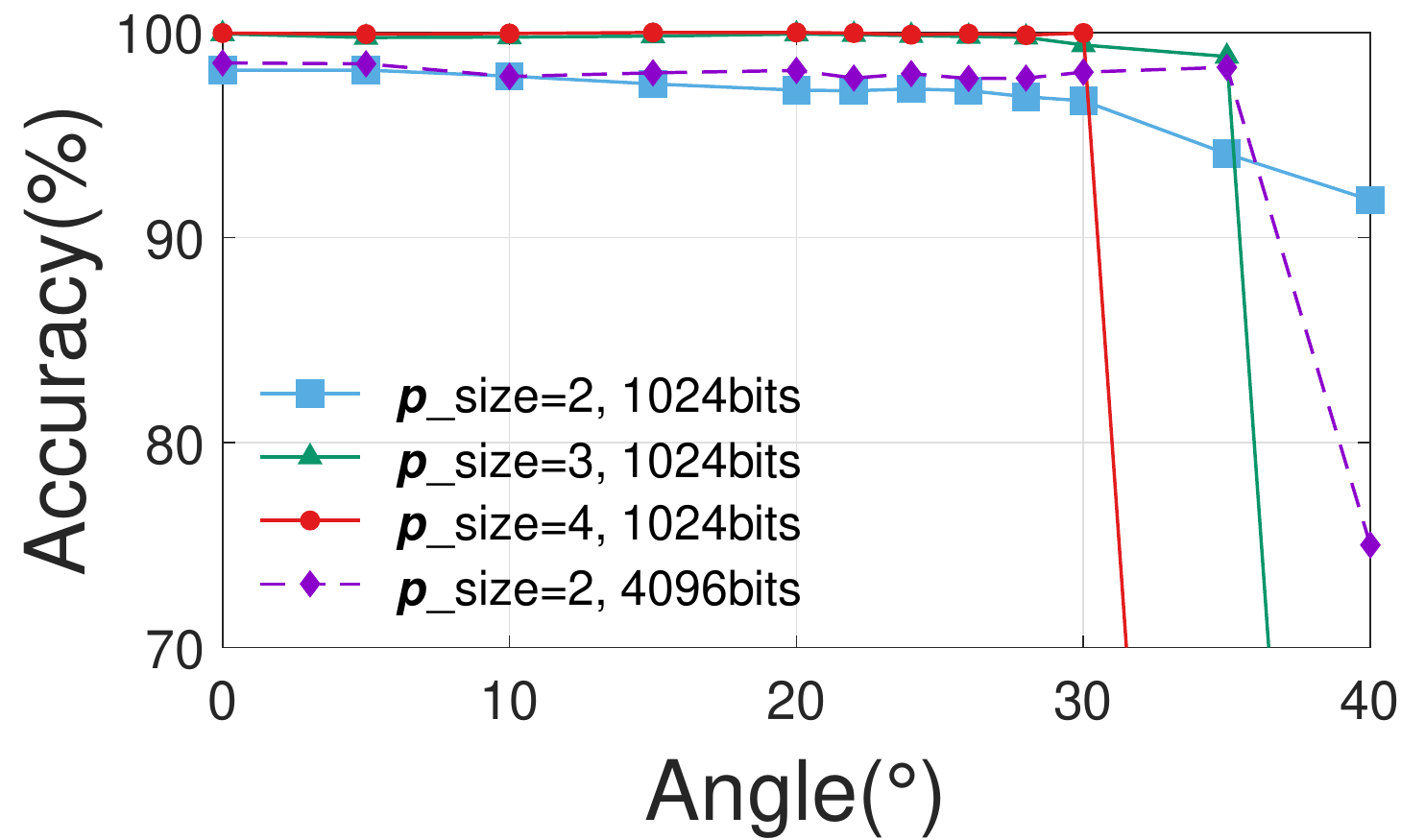}
        }
    }
    \end{center}
\caption{The detection and demodulation range of OAcode generated by different spread spectrum modules $\boldsymbol{p}$.
}
\label{psize}
\end{figure}

As shown in Fig.~\ref{psize}, blue, green, and red lines illustrate the detection and demodulation range of OAcodes with spread spectrum module size $\boldsymbol{p}\_size=2$, $3$, and $4$ pixels. When $\boldsymbol{p}\_size=4$, OAcode has the best demodulation accuracy in both distance and angle experiments ($83.13\%$ at $25cm$ and $99.98\%$ at $30^\circ$). Moreover, it has a better detection rate in the distance experiment ($40\%$ at $25cm$). \rev{These results are consistent with the conclusion mentioned in Sec.~\ref{DU_gen}, i.e., the larger size of spread spectrum module $\boldsymbol{p}$ contributes to better robustness against imaging distortions.}

On the other hand, the bigger spread spectrum module size $\boldsymbol{p}\_size$ means fewer data units in OAcode, i.e., fewer symmetrical peaks in peaks map, which increase the difficulty of OAcode synchronization. This explains why OAcode with $\boldsymbol{p}\_size=4$ has the lowest detection rate in the angle experiment, as shown in Fig.~\ref{psize} (c). On the contrary, the smaller $\boldsymbol{p}$ means better synchronization performance and lower demodulation accuracy. As shown in Fig.~\ref{psize} (c), when $\boldsymbol{p}\_size=2$, OAcode has a relatively better detection rate $30\%$ at angle $30^\circ$. Yet it has a relatively lower demodulation accuracy in Fig.~\ref{psize} (b) and (d). For the OAcode generated by $3\times 3$ size spread spectrum module, it has an intermediate detection and demodulation performance between the above two.

\subsubsection{Maximum Information Capacity}
The upper bound of the information capacity of OAcode can be determined according to the discussion of parameters in Sec.~\ref{proto_gen_section}. Specifically, there should be more than $5 \times 5$ data units in one OAcode, and the spread spectrum module $\boldsymbol{p}$ should be larger than or equal to $2 \times 2$ pixels. Thus, the maximum bit pixel rate of OAcode is $1/100$. For OAcode of $640 \times 640$ pixels in experiments, its maximum information capacity is $4096$ bits, whose performance is shown in Fig.~\ref{psize} as the purple dotted line. At an $18cm$ screen-camera distance, that OAcode has a $90\%$ detection rate and over $80\%$ demodulation accuracy. Additionally, at a $25^\circ$ shooting angle, it has a nearly $100\%$ detection rate and over $97\%$ demodulation accuracy.

\subsubsection{Extraction Time and Performance}
With the unoptimized Matlab implementation on a PC with AMD Ryzen 5 1600X CPU and NVIDIA GTX 1050 GPU, the average OAcode extraction time with fast demodulation is $1.48s$. For the enhanced demodulation, as shown in Table~\ref{time}, the number of subunits affects the average extraction time and demodulation accuracy. The enhanced demodulation with $1 \times 1$ subunit is equivalent to the fast demodulation, having the same time cost and performance, reducing $0\%$ bit error rate (BER). With the increase in the number of subunits, the enhanced demodulation reduces the BER by about $4\%$ on average but takes more time. It should be noted that the increment of the reduced BER becomes slow when the number of subunits is over $2 \times 2$. The possible reason is that $2 \times 2$ subunits recover almost all lens distortions in our camera-shooting conditions. As a result, it is appropriate to set the number of subunits of the enhanced demodulation as $2 \times 2$.

\begin{table}[t]
\large
\caption{The reduced BER and the extraction time cost of different demodulation settings.}
\begin{center}
\label{time}
\resizebox{\linewidth}{!}{
\setlength{\tabcolsep}{.1in}{
\begin{tabular}{l|ccccc}
\toprule
Number of subunits & $1 \times 1$ & $2 \times 2$ & $3\times 3$ & $4 \times 4$ \\
\midrule
Reduced BER (\%)  & 0 & 3.891 & 4.182 & 4.045  \\
\midrule
Demodulation time (s) & 1.48 & 3.13 & 5.52 & 8.64 \\ 
\bottomrule
\end{tabular}
}}
\end{center}
\end{table}

\begin{table}[h]
\large
\begin{center}
\caption{The PSNR of comparative aesthetic 2D barcode schemes,  one is calculated by their data area without position detection patterns and another one is calculated by the overall 2D barcode.}
\label{compare_psnr}
\resizebox{\linewidth}{!}{
\setlength{\tabcolsep}{.08in}{
\begin{tabular}{l|cccc}
\toprule
PSNR & PiCode & RA Code & QR images & OAcode \\
\midrule
Data area only & 24.93 & 24.98 & 25.11 & \textbf{25.12}  \\
\midrule
Overall & 15.04 & 15.98 & 10.40 & \textbf{25.12} \\
\bottomrule
\end{tabular}
}}
\end{center}
\end{table}

\begin{table}[t]
\large
\begin{center}
\caption{The extraction accuracy of OAcode and comparative aesthetic 2D barcodes under different capacities and angle settings.}
\label{comparative}
\resizebox{\linewidth}{!}{
\begin{tabular}{l|ccc|cc}
\toprule
 &  &  &  & \multicolumn{2}{c}{OAcode} \\
Accuracy (\%) & PiCode & RA Code & QR image & Fast & Enhanced \\
\midrule
$29\times29,0^\circ$ & 88.361 & 97.366 & 95.321 & 97.079 & \textbf{99.516} \\
$29\times29,20^\circ$ & 89.096 & 96.328 & 94.532 & 98.472 & \textbf{99.409} \\
\midrule
$65\times65,0^\circ$ & 94.998 & 97.686 & 90.793 & 91.667 & \textbf{99.567} \\
$65\times65,20^\circ$ & 87.051 & 96.712 & 80.430 & 89.258 & \textbf{97.887}\\
\bottomrule
\end{tabular}
}
\end{center}
\end{table}

\subsubsection{Comparative Experiment with Aesthetic 2D Barcode}
\label{sec:comparative1}
We compare the proposed OAcode with representative aesthetic 2D barcode schemes, including Picode\cite{PiCode}, RA code\cite{racode}, and QR image\cite{QRimage}. \rev{For all these methods, we use the same background images setup as Sec.~\ref{sec:overall_performance}, including $100$ images of different categories.} For a fair comparison, the setting of comparative experiments refers to \cite{PiCode,racode}. Specifically, aesthetic 2D barcodes are captured at a fixed screen-camera distance of $8cm$ with two angles $0^\circ$ or $20^\circ$. Their information capacity is set as $29\times29$ or $65\times65$ bits. Because the proposed OAcode has no position detection patterns, as shown in Table~\ref{compare_psnr}, we set the PSNR of their data area before and after beautification to the same level of $25.03\pm0.1$. It should be noted that if position detection patterns are also considered as modifications to the background image, the PSNR of overall aesthetic 2D barcodes would decrease sharply except for the proposed OAcode.

As shown in Table~\ref{comparative}, in all settings, OAcode has the highest extraction accuracy, which is higher than $99\%$. Moreover, the experiment result further proves the effectiveness of enhanced demodulation, especially in high information capacity. In the information capacity of $29 \times 29$ bits, compared to the fast demodulation, the enhanced demodulation improves the demodulation accuracy by around $2\%$. Yet in the capacity of $65 \times 65$ bits, the accuracy improved by enhanced demodulation is nearly $10\%$.

\rev{
\begin{table}[t]
\large
\begin{center}
\caption{The extraction accuracy of OAcode and comparative deep data hiding methods under different angle settings.}
\label{tab:comparative2}
\resizebox{\linewidth}{!}{
\setlength{\tabcolsep}{.2in}{
\begin{tabular}{l|ccc}
\toprule
 Methods & Hidden Code & StegaStamp & OAcode \\
\midrule
PSNR & 34.035 & 34.011 & \textbf{34.181} \\
\midrule
Acc (\%), $0^\circ$ & 90.327 & 92.480 & \textbf{93.245} \\
\midrule
Acc (\%), $20^\circ$ & 87.122 & 90.770 & \textbf{91.204} \\
\bottomrule
\end{tabular}
}}
\end{center}
\end{table}
\subsubsection{Comparative Experiment with Deep Data Hiding}
In state-of-art deep data hiding methods, some of them have similar application scenarios to 2D barcodes. Thus, we select two representative methods with available evaluation models, including StegaStamp \cite{2019stegastamp} and Hidden Code \cite{jia2022learning} to compare with OAcode. Since we only have the pre-trained evaluation model of Hidden Code, whose parameters are hard to modify, we adjust the OAcode and StegaStamp to align with Hidden Code for fair comparative experiments. Specifically, using these methods, we embed $196$ bits in background images of $256\times 256$ pixels and set their PSNR to the same level of $34.1\pm0.1$. The image dataset and capturing setting are the same as Sec.~\ref{sec:comparative1}. As shown in Table~\ref{tab:comparative2}, under the two common camera angles, OAcode has the highest extraction accuracy, higher than $90\%$. One possible reason is that OAcode deals with lens distortion better with the enhanced demodulation. Additionally, we notice that Hidden Code and StegaStamp have relatively low extraction accuracy on logo images, probably because they are originally designed for natural images.
}

\begin{figure}[t]
\setlength{\belowcaptionskip}{-5pt}
\setlength{\abovecaptionskip}{0pt}
  \begin{center}
  \includegraphics[width=3.4in]{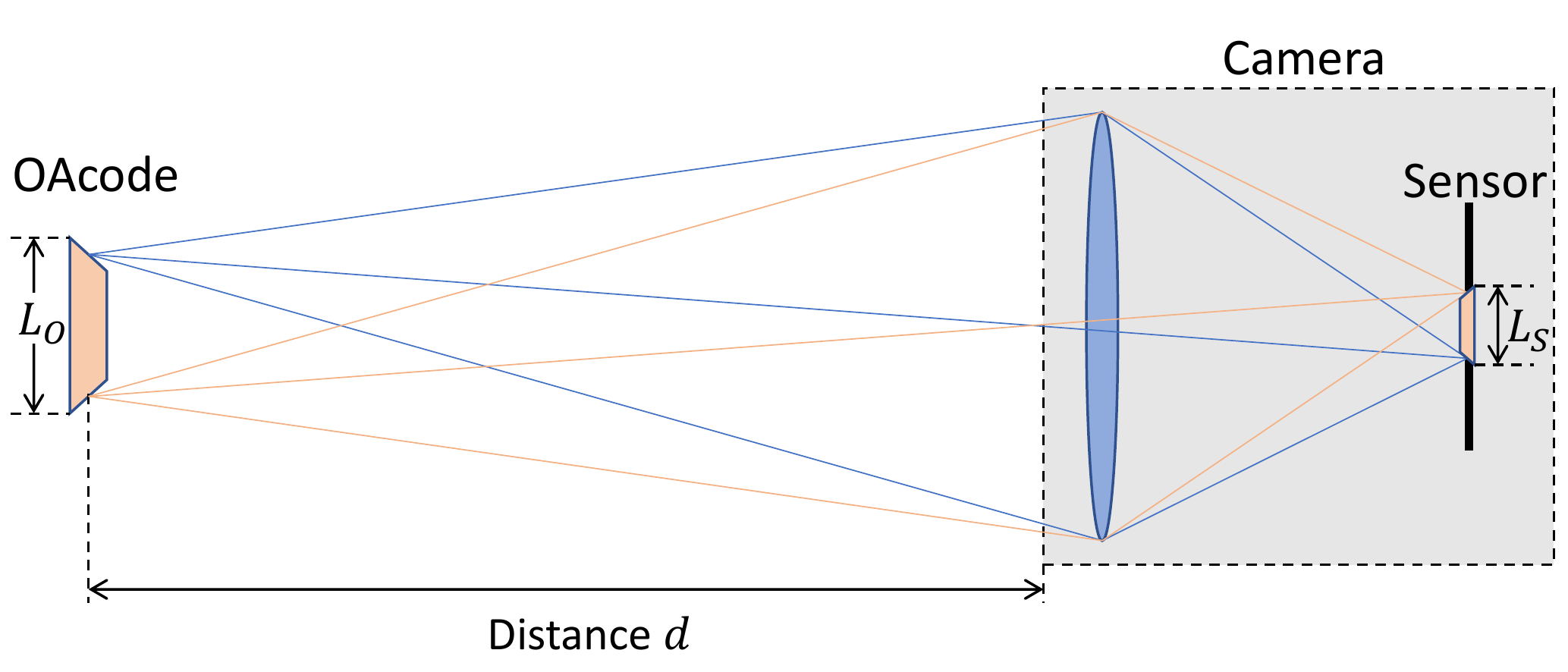}
  \caption{Schematic diagram of the camera-shooting process of OAcode.}
  \label{shooting_model}
  \end{center}
\end{figure}

\subsubsection{Performance on Other Devices or Parameters}

One of the benefits of using an experimental setup with explicitly defined parameters is that for new devices or parameters, we can roughly estimate the performance of OAcode. Specifically, the camera-shooting process can be modeled as Fig.~\ref{shooting_model}, where an OAcode with physical side length $L_O$ is captured by the camera at distance $d$, and converted into an image with side length $L_S$ on the camera sensor with resolution $r$. In this model, the final captured pixel size $L_p$ of OAcode, which considerably affects the OAcode extraction result, has the following relationship with the above parameters roughly:

\begin{equation}
\label{dis_relation}
    L_p \propto L_S \cdot r \propto \frac{L_O \cdot r}{d},
\end{equation}
which could be used to estimate the performance of OAcode under the new parameters. For example, according to the experimental result in Sec.~\ref{extraction_range}, OAcode with side length $L_O=5cm$ has $100\%$ detection rate and $96.76\%$ demodulation accuracy under the camera-shooting environment with distance $d=15cm$ and camera resolution $r=1280\times 720$ pixels. According to Eq.~(\ref{dis_relation}), when OAcode is displayed with double size $L_O=10cm$ and the camera resolution is about three times larger 720p, OAcode is supposed to have similar extraction performance at $2\times 3=6$ times the distance, i.e., $15\times 6=90cm$. We use different brands of phones with default camera resolutions to verify this assumption. As shown in Table~\ref{more_dis}, OAcode has a longer available distance under better devices or parameters and roughly conforms to the relationship in Eq.~(\ref{dis_relation}).

\begin{table}[h]
\large
\begin{center}
\caption{The detection rate and demodulation accuracy of OAcode captured by different phones at a distance of $90cm$.}
\label{more_dis}
\resizebox{\linewidth}{!}{
\setlength{\tabcolsep}{.07in}{
\begin{tabular}{l|cccc}
\toprule
Phones & Xiaomi Mi6 & Redmi K40 & Huawei P30 Pro &  iPhone 11 \\
\midrule
Resolution & $3016\times 4032$ & \rev{$3000 \times 3000$} & $2976 \times 2976$ &  $4032 \times 2268$ \\
\midrule
Det rate (\%) & 80 & 100 & 100 &  100 \\
\midrule
Acc (\%) & 89.355 & 94.785 & 93.681 & 97.100 \\
\bottomrule
\end{tabular}
}}
\end{center}
\end{table}

\section{Conclusion}
In this paper, we propose a novel overall aesthetic 2D barcode scheme, named OAcode, which has no position detection patterns but achieves all functions that a complete aesthetic 2D barcode should have. Experimental results illustrate that, at certain screen-camera distances and angles, OAcode could be quickly detected and demodulated. Additionally, the performance of OAcode could be further improved with better devices and parameters. In our future work, we will explore the application of OAcode in document printing and investigate a crop-resistant 2D barcode based on the framework of OAcode.

\bibliographystyle{IEEEtran}
\bibliography{IEEEabrv,main}

\begin{thebibliography}{10}
\providecommand{\url}[1]{#1}
\csname url@rmstyle\endcsname
\providecommand{\newblock}{\relax}
\providecommand{\bibinfo}[2]{#2}
\providecommand\BIBentrySTDinterwordspacing{\spaceskip=0pt\relax}
\providecommand\BIBentryALTinterwordstretchfactor{4}
\providecommand\BIBentryALTinterwordspacing{\spaceskip=\fontdimen2\font plus
\BIBentryALTinterwordstretchfactor\fontdimen3\font minus
  \fontdimen4\font\relax}
\providecommand\BIBforeignlanguage[2]{{%
\expandafter\ifx\csname l@#1\endcsname\relax
\typeout{** WARNING: IEEEtran.bst: No hyphenation pattern has been}%
\typeout{** loaded for the language `#1'. Using the pattern for}%
\typeout{** the default language instead.}%
\else
\language=\csname l@#1\endcsname
\fi
#2}}

\bibitem{device}
E.~{Ohbuchi}, H.~{Hanaizumi}, and L.~A. {Hock}, ``Barcode readers using the
  camera device in mobile phones,'' in \emph{2004 International Conference on
  Cyberworlds}, 2004, pp. 260--265.

\bibitem{qr_mobile_marketing}
\BIBentryALTinterwordspacing
M.~S. Hossain, X.~Zhou, and M.~F. Rahman, ``Examining the impact of qr codes on
  purchase intention and customer satisfaction on the basis of perceived
  flow,'' \emph{International Journal of Engineering Business Management},
  vol.~10, p. 1847979018812323, 2018. [Online]. Available:
  \url{https://doi.org/10.1177/1847979018812323}
\BIBentrySTDinterwordspacing

\bibitem{cata2013qr}
T.~Cata, P.~S. Patel, and T.~Sakaguchi, ``Qr code: A new opportunity for
  effective mobile marketing,'' \emph{Journal of Mobile technologies, knowledge
  and society}, vol. 2013, p.~1, 2013.

\bibitem{okazaki2012benchmarking}
S.~Okazaki, H.~Li, and M.~Hirose, ``Benchmarking the use of qr code in mobile
  promotion: three studies in japan,'' \emph{Journal of Advertising Research},
  vol.~52, no.~1, pp. 102--117, 2012.

\bibitem{usage_in_2022}
\BIBentryALTinterwordspacing
``Research on qr code marketing: Effectiveness, adoption, and use cases,''
  2022. [Online]. Available:
  \url{https://www.qrcodechimp.com/qr-code-marketing-research/}
\BIBentrySTDinterwordspacing

\bibitem{min2021screen}
X.~Min, K.~Gu, G.~Zhai, X.~Yang, W.~Zhang, P.~Le~Callet, and C.~W. Chen,
  ``Screen content quality assessment: overview, benchmark, and beyond,''
  \emph{ACM Computing Surveys (CSUR)}, vol.~54, no.~9, pp. 1--36, 2021.

\bibitem{ma2022learning}
K.~Ma, S.~Das, Z.~Shu, and D.~Samaras, ``Learning from documents in the wild to
  improve document unwarping,'' in \emph{ACM SIGGRAPH 2022 Conference
  Proceedings}, 2022, pp. 1--9.

\bibitem{sellen2003myth}
A.~J. Sellen and R.~H. Harper, \emph{The myth of the paperless office}.\hskip
  1em plus 0.5em minus 0.4em\relax MIT press, 2003.

\bibitem{QRimage}
G.~J. {Garateguy}, G.~R. {Arce}, D.~L. {Lau}, and O.~P. {Villarreal}, ``Qr
  images: Optimized image embedding in qr codes,'' \emph{IEEE Transactions on
  Image Processing}, vol.~23, no.~7, pp. 2842--2853, 2014.

\bibitem{PiCode}
W.~{Huang} and W.~H. {Mow}, ``Picode: 2d barcode with embedded picture and
  vicode: 3d barcode with embedded video,'' in \emph{Proceedings of the 19th
  annual international conference on Mobile computing \& networking}, 2013, pp.
  139--142.

\bibitem{racode}
C.~{Chen}, B.~{Zhou}, and W.~H. {Mow}, ``Ra code: A robust and aesthetic code
  for resolution-constrained applications,'' \emph{IEEE Transactions on
  Circuits and Systems for Video Technology}, vol.~28, no.~11, pp. 3300--3312,
  2018.

\bibitem{rucode}
C.~Chen, W.~Huang, L.~Zhang, and W.~H. Mow, ``Robust and unobtrusive
  display-to-camera communications via blue channel embedding,'' \emph{IEEE
  Transactions on Image Processing}, vol.~28, no.~1, pp. 156--169, 2019.

\bibitem{qrcode}
\BIBentryALTinterwordspacing
``Qrcode,'' 2020. [Online]. Available:
  \url{https://www.qrcode.com/zh/index.html}
\BIBentrySTDinterwordspacing

\bibitem{datamatrix}
\BIBentryALTinterwordspacing
``Data matrix,'' 2020. [Online]. Available:
  \url{https://en.wikipedia.org/wiki/Data_Matrix}
\BIBentrySTDinterwordspacing

\bibitem{zhai2020perceptual}
G.~Zhai and X.~Min, ``Perceptual image quality assessment: a survey,''
  \emph{Science China Information Sciences}, vol.~63, no.~11, pp. 1--52, 2020.

\bibitem{min2017blind}
X.~Min, K.~Gu, G.~Zhai, J.~Liu, X.~Yang, and C.~W. Chen, ``Blind quality
  assessment based on pseudo-reference image,'' \emph{IEEE Transactions on
  Multimedia}, vol.~20, no.~8, pp. 2049--2062, 2017.

\bibitem{min2018blind}
X.~Min, G.~Zhai, K.~Gu, Y.~Liu, and X.~Yang, ``Blind image quality estimation
  via distortion aggravation,'' \emph{IEEE Transactions on Broadcasting},
  vol.~64, no.~2, pp. 508--517, 2018.

\bibitem{min2020study}
X.~Min, G.~Zhai, J.~Zhou, M.~C. Farias, and A.~C. Bovik, ``Study of subjective
  and objective quality assessment of audio-visual signals,'' \emph{IEEE
  Transactions on Image Processing}, vol.~29, pp. 6054--6068, 2020.

\bibitem{min2017unified}
X.~Min, K.~Ma, K.~Gu, G.~Zhai, Z.~Wang, and W.~Lin, ``Unified blind quality
  assessment of compressed natural, graphic, and screen content images,''
  \emph{IEEE Transactions on Image Processing}, vol.~26, no.~11, pp.
  5462--5474, 2017.

\bibitem{min2018saliency}
X.~Min, K.~Gu, G.~Zhai, M.~Hu, and X.~Yang, ``Saliency-induced
  reduced-reference quality index for natural scene and screen content
  images,'' \emph{Signal Processing}, vol. 145, pp. 127--136, 2018.

\bibitem{qart}
\BIBentryALTinterwordspacing
R.~Cox, ``Qartcodes,'' April 2012. [Online]. Available:
  \url{http://research.swtch.com/qart}
\BIBentrySTDinterwordspacing

\bibitem{chu2013halftone}
H.-K. Chu, C.-S. Chang, R.-R. Lee, and N.~J. Mitra, ``Halftone qr codes,''
  \emph{ACM Transactions on Graphics (TOG)}, vol.~32, no.~6, pp. 1--8, 2013.

\bibitem{Eff_Beau}
S.~{Lin}, M.~{Hu}, C.~{Lee}, and T.~{Lee}, ``Efficient qr code beautification
  with high quality visual content,'' \emph{IEEE Transactions on Multimedia},
  vol.~17, no.~9, pp. 1515--1524, 2015.

\bibitem{Visualead}
\BIBentryALTinterwordspacing
N.~{Aliva}, U.~{Peled}, and F.~{Itamar}, ``Visualead,'' 2020. [Online].
  Available: \url{https://www.visualead.com/}
\BIBentrySTDinterwordspacing

\bibitem{Visually}
Z.~{Baharav} and R.~{Kakarala}, ``Visually significant qr codes: Image blending
  and statistical analysis,'' in \emph{2013 IEEE International Conference on
  Multimedia and Expo (ICME)}, 2013, pp. 1--6.

\bibitem{appearance}
Y.~{Lin}, Y.~{Chang}, and J.~{Wu}, ``Appearance-based qr code beautifier,''
  \emph{IEEE Transactions on Multimedia}, vol.~15, no.~8, pp. 2198--2207, 2013.

\bibitem{Stylized}
M.~{Xu}, H.~{Su}, Y.~{Li}, X.~{Li}, J.~{Liao}, J.~{Niu}, P.~{Lv}, and
  B.~{Zhou}, ``Stylized aesthetic qr code,'' \emph{IEEE Transactions on
  Multimedia}, vol.~21, no.~8, pp. 1960--1970, 2019.

\bibitem{yang2016artcode}
Z.~Yang, Y.~Bao, C.~Luo, X.~Zhao, S.~Zhu, C.~Peng, Y.~Liu, and X.~Wang,
  ``Artcode: preserve art and code in any image,'' in \emph{Proceedings of the
  2016 ACM International Joint Conference on Pervasive and Ubiquitous
  Computing}, 2016, pp. 904--915.

\bibitem{liu2011toward}
J.-C. Liu and H.-A. Shieh, ``Toward a two-dimensional barcode with visual
  information using perceptual shaping watermarking in mobile applications,''
  \emph{Optical Engineering}, vol.~50, no.~1, p. 017002, 2011.

\bibitem{EMBDN}
J.~Jia, G.~Zhai, J.~Zhang, Z.~Gao, Z.~Zhu, X.~Min, X.~Yang, and G.~Guo,
  ``Embdn: An efficient multiclass barcode detection network for complicated
  environments,'' \emph{IEEE Internet of Things Journal}, vol.~6, no.~6, pp.
  9919--9933, 2019.

\bibitem{Tiny-BDN}
J.~Jia, G.~Zhai, P.~Ren, J.~Zhang, Z.~Gao, X.~Min, and X.~Yang, ``Tiny-bdn: An
  efficient and compact barcode detection network,'' \emph{IEEE Journal of
  Selected Topics in Signal Processing}, vol.~14, no.~4, pp. 688--699, 2020.

\bibitem{ZhangMJZWZ21}
\BIBentryALTinterwordspacing
J.~Zhang, X.~Min, J.~Jia, Z.~Zhu, J.~Wang, and G.~Zhai, ``Fine localization and
  distortion resistant detection of multi-class barcode in complex
  environments,'' \emph{Multim. Tools Appl.}, vol.~80, no.~11, pp.
  16\,153--16\,172, 2021. [Online]. Available:
  \url{https://doi.org/10.1007/s11042-019-08578-x}
\BIBentrySTDinterwordspacing

\bibitem{qr_area}
\BIBentryALTinterwordspacing
D.~W. {Inc.} and {Elkhart}, ``Securing margin,'' 2020. [Online]. Available:
  \url{https://www.qrcode.com/en/howto/code.html}
\BIBentrySTDinterwordspacing

\bibitem{lens_distortion}
S.~Kang, S.~D. Kim, and M.~Kim, ``Structural-information-based robust corner
  point extraction for camera calibration under lens distortions and
  compression artifacts,'' \emph{IEEE Access}, vol.~9, pp. 151\,037--151\,048,
  2021.

\bibitem{stirmark1}
F.~A. Petitcolas, R.~J. Anderson, and M.~G. Kuhn, ``Attacks on copyright
  marking systems,'' in \emph{International workshop on information
  hiding}.\hskip 1em plus 0.5em minus 0.4em\relax Springer, 1998, pp. 218--238.

\bibitem{stirmark2}
F.~A. Petitcolas, ``Watermarking schemes evaluation,'' \emph{IEEE Signal
  Process. Mag}, vol.~17, no.~5, pp. 58--64, 2000.

\bibitem{hiding}
F.~A.~P. {Petitcolas}, R.~J. {Anderson}, and M.~G. {Kuhn}, ``Information
  hiding-a survey,'' \emph{Proceedings of the IEEE}, vol.~87, no.~7, pp.
  1062--1078, 1999.

\bibitem{history2}
C.~I. Podilchuk and E.~J. Delp, ``Digital watermarking: algorithms and
  applications,'' \emph{IEEE Signal Process. Mag}, vol.~18, no.~4, pp. 33--46,
  2001.

\bibitem{fang2019camera}
H.~Fang, W.~Zhang, Z.~Ma, H.~Zhou, S.~Sun, H.~Cui, and N.~Yu, ``A camera
  shooting resilient watermarking scheme for underpainting documents,''
  \emph{IEEE Trans. Circuits Syst. Video Technol}, 2019.

\bibitem{Fanghan}
H.~Fang, W.~Zhang, H.~Zhou, H.~Cui, and N.~Yu, ``Screen-shooting resilient
  watermarking,'' \emph{IEEE Trans. Inf. Forensics Security}, vol.~14, no.~6,
  pp. 1403--1418, 2018.

\bibitem{steganography1}
\BIBentryALTinterwordspacing
A.~Cheddad, J.~Condell, K.~Curran, and P.~{Mc Kevitt}, ``Digital image
  steganography: Survey and analysis of current methods,'' \emph{Signal
  Processing}, vol.~90, no.~3, pp. 727--752, 2010. [Online]. Available:
  \url{https://www.sciencedirect.com/science/article/pii/S0165168409003648}
\BIBentrySTDinterwordspacing

\bibitem{steganography2}
N.~Provos and P.~Honeyman, ``Hide and seek: an introduction to steganography,''
  \emph{IEEE Security \& Privacy}, vol.~1, no.~3, pp. 32--44, 2003.

\bibitem{steganography3}
N.~Subramanian, O.~Elharrouss, S.~Al-Maadeed, and A.~Bouridane, ``Image
  steganography: A review of the recent advances,'' \emph{IEEE Access}, vol.~9,
  pp. 23\,409--23\,423, 2021.

\bibitem{2019stegastamp}
M.~Tancik, B.~Mildenhall, and R.~Ng, ``Stegastamp: Invisible hyperlinks in
  physical photographs,'' in \emph{IEEE Conference on Computer Vision and
  Pattern Recognition (CVPR)}, 2020.

\bibitem{TERA}
H.~Fang, D.~Chen, F.~Wang, Z.~Ma, H.~Liu, W.~Zhou, W.~Zhang, and N.~Yu, ``Tera:
  Screen-to-camera image code with transparency, efficiency, robustness and
  adaptability,'' \emph{IEEE Transactions on Multimedia}, vol.~24, pp.
  955--967, 2022.

\bibitem{RIHOOP}
J.~Jia, Z.~Gao, K.~Chen, M.~Hu, X.~Min, G.~Zhai, and X.~Yang, ``Rihoop: Robust
  invisible hyperlinks in offline and online photographs,'' \emph{IEEE
  Transactions on Cybernetics}, vol.~52, no.~7, pp. 7094--7106, 2022.

\bibitem{jia2022learning}
J.~Jia, Z.~Gao, D.~Zhu, X.~Min, G.~Zhai, and X.~Yang, ``Learning invisible
  markers for hidden codes in offline-to-online photography,'' in
  \emph{Proceedings of the IEEE/CVF Conference on Computer Vision and Pattern
  Recognition}, 2022, pp. 2273--2282.

\bibitem{loc_distortion}
Z.~Ma, W.~Zhang, H.~Fang, X.~Dong, L.~Geng, and N.~Yu, ``Local geometric
  distortions resilient watermarking scheme based on symmetry,'' \emph{IEEE
  Transactions on Circuits and Systems for Video Technology}, pp. 1--1, 2021.

\bibitem{Theory_SS}
R.~Pickholtz, D.~Schilling, and L.~Milstein, ``Theory of spread-spectrum
  communications - a tutorial,'' \emph{IEEE Transactions on Communications},
  vol.~30, no.~5, pp. 855--884, 1982.

\bibitem{CHAPMAN201535}
\BIBentryALTinterwordspacing
T.~Chapman, E.~Larsson, P.~{von Wrycza}, E.~Dahlman, S.~Parkvall, and
  J.~Sköld, ``Chapter 3 - cdma transmission principles,'' in \emph{HSPA
  Evolution}.\hskip 1em plus 0.5em minus 0.4em\relax Oxford: Academic Press,
  2015, pp. 35--48. [Online]. Available:
  \url{https://www.sciencedirect.com/science/article/pii/B978008099969200003X}
\BIBentrySTDinterwordspacing

\bibitem{book_wiener}
J.~S. Lim, \emph{Two-Dimensional Signal and Image Processing}.\hskip 1em plus
  0.5em minus 0.4em\relax Englewood Cliffs, NJ, Prentice Hall., 1990, p. 548,
  equations 9.26, 9.27, and 9.29.

\bibitem{BelgaLogos}
\BIBentryALTinterwordspacing
``Belgalogos dataset,'' March 2014. [Online]. Available:
  \url{http://www-sop.inria.fr/members/Alexis.Joly/BelgaLogos/BelgaLogos.html}
\BIBentrySTDinterwordspacing

\bibitem{CVG}
\BIBentryALTinterwordspacing
``Cvg-ugr image database,'' March 2014. [Online]. Available:
  \url{http://decsai.ugr.es/cvg/dbimagenes/}
\BIBentrySTDinterwordspacing

\end{thebibliography}

\vfill
\end{document}